\documentclass{ieeeaccess}
\usepackage{cite}
\usepackage{amsmath,amssymb,amsfonts}
\usepackage{graphicx}
\usepackage{textcomp,multirow,url,tabularx,stfloats}

\usepackage{bm}
\makeatletter
\AtBeginDocument{\DeclareMathVersion{bold}
\SetSymbolFont{operators}{bold}{T1}{times}{b}{n}
\SetSymbolFont{NewLetters}{bold}{T1}{times}{b}{it}
\SetMathAlphabet{\mathrm}{bold}{T1}{times}{b}{n}
\SetMathAlphabet{\mathit}{bold}{T1}{times}{b}{it}
\SetMathAlphabet{\mathbf}{bold}{T1}{times}{b}{n}
\SetMathAlphabet{\mathtt}{bold}{OT1}{pcr}{b}{n}
\SetSymbolFont{symbols}{bold}{OMS}{cmsy}{b}{n}
\renewcommand\boldmath{\@nomath\boldmath\mathversion{bold}}}
\makeatother

\def\BibTeX{{\rm B\kern-.05em{\sc i\kern-.025em b}\kern-.08em
    T\kern-.1667em\lower.7ex\hbox{E}\kern-.125emX}}

\makeatletter
\def\submissionfooter{\hbox to\textwidth{\footervolfont Submitted manuscript\hfil\footerpagefont\thepage}}
\g@addto@macro\ps@titlepage{\let\@oddfoot\submissionfooter\let\@evenfoot\submissionfooter}
\g@addto@macro\ps@headings{\let\@oddfoot\submissionfooter\let\@evenfoot\submissionfooter}
\makeatother
\usepackage[hidelinks]{hyperref}
\begin{document}
\history{Manuscript submitted for review.}
\doi{To be assigned}
\title{Digital-Twin-Referenced Belief-Guided Q-Learning for Jammer Identification and Beam-Power Control in a mmWave Model}
\author{\uppercase{Parmida Geranmayeh}\authorrefmark{1} and \uppercase{Onur G\"unl\"u}\authorrefmark{1,2}}
\address[1]{Lehrstuhl f\"ur Nachrichtentechnik, Technische Universit\"at Dortmund, 44227 Dortmund, Germany (e-mail: parmida.geranmayeh@tu-dortmund.de; onur.guenlue@tu-dortmund.de)}
\address[2]{Information Theory and Security Laboratory (ITSL), Link\"oping University, 581 83 Link\"oping, Sweden}
\markboth{Geranmayeh and G\"unl\"u: DT-Referenced BGQL for Jammer Identification}{Geranmayeh and G\"unl\"u: DT-Referenced BGQL for Jammer Identification}
\corresp{Corresponding author: Parmida Geranmayeh (e-mail: parmida.geranmayeh@tu-dortmund.de).}
\begin{abstract}
Next-generation wireless networks are expected to support high data rates and reliable connectivity while meeting demanding requirements for energy efficiency, resource management, and communication security. Digital Twin (DT) technology provides a basis for addressing these requirements by representing expected network behavior and enabling the actual network state to be continuously assessed against it. Integrating DT technology with reinforcement learning (RL) further supports adaptive resource allocation and security-aware decision-making in dynamic wireless environments. This paper presents a simulation-based digital-twin (DT) reference and a belief-guided Q-learning (BGQL) method for joint jammer identification and beam--power control in a site-specific 28-GHz network model. For each action, the same channel simulator is evaluated under an attack-free reference configuration and an attacked configuration. Receiver-level residuals are combined with candidate spatial jamming footprints and accumulated over time to form a normalized suspicion vector. BGQL incorporates this vector into an aggregated state, a belief-concentration reward term, and its exploratory beam-selection rule. The complete BGQL scheme is compared with SARSA, Greedy, and Random under matched attacker sets. Across 30 paired Monte Carlo runs in each of four reported configurations, with the number of active jammers supplied to every method, BGQL achieves mean aggregate full-band link-rate scores of approximately 437.4--564.2~Mbit/s and rate-to-device-power scores of 60.37--76.79~Mbit/J. Its top-$N_{\mathrm{Att}}$ localization recall is $100\%$ in the one-jammer configurations and approximately $78\%$--$81\%$ in the two-jammer configurations. BGQL requires approximately 80--130~s per run, compared with 5--14~s for SARSA, reflecting its evaluation of $25^2=625$ candidate beam pairs during exploratory steps. These findings provide a matched-model simulation proof of concept for DT-referenced belief-guided learning for security-aware beam-power control in 6G networks.
\end{abstract}

\begin{keywords}
Belief-guided Q-learning ,
Beamforming optimization ,
Digital-twin residuals ,
Energy-aware beam-power control ,
Jammer identification
\end{keywords}

\maketitle
\pagestyle{headings}

\section{Introduction}
\label{sec:introduction}

Next-generation wireless networks face a set of simultaneous and sometimes conflicting requirements. On the one hand, the increasing demand for higher data rates and more efficient utilization of limited spectrum resources has increased the need for techniques such as spatial reuse and multiplexing. On the other hand, increasing user density, dynamic channel conditions, energy constraints, and the emergence of intelligent threats have created a growing need for adaptive resource management and advanced security mechanisms. In response to these requirements, sixth-generation (6G) networks are expected to extensively utilize artificial intelligence (AI) and machine learning (ML) to achieve higher levels of automation, reliability, and operational efficiency \cite{Fong2023Reliability,Haider2020AI}. Applications of ML in communication systems cover a wide range of problems, including source and channel coding \cite{Bourtsoulatze2019DeepJSCC,Nachmani2018Decoding,Kim2020PhysicalLayer}, waveform design \cite{Aoudia2022Waveform}, signal detection \cite{Ye2018OFDM,Farsad2018Detection}, distributed compression \cite{Sohrabi2022Compression}, resource allocation \cite{Sun2018Optimize,Cui2019Spatial,Lee2021GraphEmbedding,Shen2021GNN,Sohrabi2021Beamforming,Zhang2022RIS}, and channel estimation \cite{He2018Channel,Neumann2018MMSE}. These advances demonstrate that data-driven methods can provide efficient and adaptive solutions to problems that become increasingly difficult to solve analytically or through exhaustive search as network dimensions increase \cite{Yu2022DeepLearning}.

In this context, the integration of sensing and communication capabilities has emerged as an important paradigm for future wireless networks \cite{OnurRoleofISAC,OnurFurkanISACTutorial}. Integrated Sensing and Communication (ISAC), which jointly utilizes infrastructure, spectrum, and processing resources for communication and sensing purposes, is recognized as one of the key enabling technologies for 6G. In related 6G research, digital twins (DTs) have been studied for reflective intelligent surface (RIS)-assisted communication \cite{Sheen2020DigitalTwinRIS}, while power-efficient antenna switching and beamforming have been investigated for multi-user SWIPT \cite{Jalali2023SWIPT}. Furthermore, the role of machine learning in joint sensing and communication, sensing-assisted communication, and communication-assisted sensing has been investigated \cite{Demirhan2023ISAC}. RL-based methods have also been proposed for radar activity prediction and the identification of channel access opportunities \cite{Salami2023Radar}. The development of dedicated AI/ML models for practical network applications has been emphasized in \cite{Ayed2023Accordion}, while \cite{Ferreira2019Satellite} investigates the potential of neural-network-based RL algorithms in different aspects of communication links. These developments indicate that network intelligence is not limited to a single task but can extend from environmental modeling to resource management and security-aware decision-making.

Among resource management mechanisms, beamforming and power control play an important role in the efficient utilization of spatial and energy resources. Beamforming enables directional transmission toward intended users, improving link quality and communication capacity while mitigating the effects of unwanted and adversarial interference \cite{Bakhtiar2023Beamforming,Chary2024AIBeamforming,Geranmayeh2022BeamSelection}. Power control also directly affects link quality, interference, and energy consumption; therefore, appropriate beam selection and transmit-power control represent two closely coupled decisions in the design of efficient wireless networks. In this context, Ihsan et al. \cite{Ihsan2022IRS} combined intelligent reflective surfaces (IRS) with non-orthogonal multiple access (NOMA)-based beamforming to jointly optimize beamforming and power allocation for improved energy efficiency. Jiang et al. \cite{Jiang2022ICC} proposed an initial beamforming strategy based on complementary beams to achieve more uniform spatial coverage. Qi et al. \cite{Qi2021UAVBeamforming} proposed a UAV- and IRS-assisted architecture for IoT applications and demonstrated that beamforming can improve the range and performance of backscatter communications. Similar results have also been reported in \cite{Qi2021UAVBeamforming,Ihsan2022IRS, Jiang2022ICC} for the application of IRS, NOMA, and complementary beams in 6G networks.

However, effective beam and power decision-making requires an accurate representation of the propagation environment. In practical wireless networks, multipath propagation, reflection, diffraction, blockage, fading, and variations in environmental geometry create a complex relationship between beam direction, transmit power, and link quality. Consequently, ray-tracing has become an important tool for realistic channel modeling and wave propagation analysis \cite{Geok2018RayTracing,Ji2021RayTracingIndoor,Hsiao2017RayTracing}. By reconstructing the dominant propagation paths and accounting for phenomena such as reflection and diffraction, ray-tracing enables more accurate characterization of the wireless channel and, consequently, a more realistic evaluation of beamforming strategies.

The importance of realistic propagation modeling has been demonstrated in various studies. Ji et al. \cite{Ji2021RayTracingIndoor} improved the computational efficiency of ray tracing for indoor UHF propagation prediction by partitioning the environment into irregular cells. Hsiao et al. \cite{Hsiao2017RayTracing} investigated millimeter-wave propagation under both line-of-sight (LOS) and non-line-of-sight (NLOS) conditions. A comprehensive review of ray-tracing algorithms and their applications to wireless channel modeling was presented in \cite{Yun2015RayTracingTheory}, while Tiberi et al. \cite{Tiberi2009UWB} investigated the characteristics of ultra-wideband channels using ray-tracing simulations. In addition to the propagation environment, realistic antenna characteristics, including radiation patterns, gain, polarization, and frequency response, directly affect the received signal energy and, consequently, beam selection \cite{Jalali2023SWIPT,Bakhtiar2023Beamforming}. Therefore, combining realistic propagation and antenna models provides a reliable basis for network-state assessment and beam and power optimization.

Beyond performance optimization, an accurate physical channel model also enables the establishment of reference network behavior. As network complexity increases, the use of such models within DT frameworks has become increasingly important \cite{Ding2022DigitalTwin,Krause2023DTAnomaly,Ruah2023BayesianDT}. A DT can represent the expected behavior of the network under normal operating conditions, enabling comparison between the attacked simulation state and its attack-free reference state. In such a framework, quantities such as the aggregate full-band link-rate score, SINR, and interference characteristics under attack-free conditions can serve as a basis for evaluating attacked simulation behavior. Therefore, deviations between expected and observed behavior can provide information not only about performance degradation but also about potential anomalies or security threats.

This capability becomes particularly important in the presence of malicious users and jamming attacks. Network security, alongside communication efficiency, has become a fundamental requirement of next-generation wireless systems \cite{OnurSecureISACJournal,OnurSecurityTutorial,Zhou2026FeedbackLunch,Onur2026secureISACTutorial}. An attacker can cause the network behavior to deviate from its expected state by generating interference, altering the received signal characteristics, or reducing SINR and the aggregate full-band link-rate score. However, detecting such deviations alone is insufficient to identify an attacker, because factors such as fading, multipath propagation, noise, and temporal channel variations can also produce similar changes in network performance \cite{Yu2023Covert,Esmaili2026Covert}. Consequently, an instantaneous decrease in the link-rate score or SINR does not necessarily indicate the presence of an attacker, and security decisions based on a single observation may lead to incorrect detection.

Reliable attacker detection therefore requires information from multiple sources and the accumulation of evidence over time. Machine-learning methods have also been investigated for wireless covert-channel detection \cite{Elsadig2022Covert}. Moreover, inference and belief-based methods can aggregate evidence from successive observations and gradually update the system's confidence regarding the likelihood that individual nodes are malicious. Bayesian inference for detecting jamming and subtle interference patterns in 5G NR networks has been investigated in \cite{Jere2023BayesianJamming}, while trust-based detection approaches incorporating active learning and Bayesian classifiers have demonstrated promising capabilities for detecting malicious behavior \cite{Huang2022VANET}. Therefore, combining deviations between expected and observed network behavior with temporal evidence aggregation can provide a more reliable basis for identifying suspicious nodes.

In addition to detection, the association of users with network resources in the presence of potentially malicious nodes represents a decision-making problem under uncertainty. Game theory provides a suitable framework for modeling such strategic interactions. Lei et al. \cite{Lei2018Game} proposed a Markov game with incomplete information to model the dynamic evolution of attack and defense strategies. Jiang et al. \cite{Jiang2021Game} proposed a multi-stage signaling game for Moving Target Defense (MTD), while Teófilo et al. \cite{Teofilo2012Poker} demonstrated the importance of adaptive decision-making under uncertainty. In related research, Ren et al. \cite{Ren2023UAV} combined incomplete-information games with DRL to derive Bayesian Nash equilibrium strategies for UAV networks, while Yao et al. \cite{Yao2024BayesianGame} proposed a Bayesian Q-learning approach for cyber defense in cyber-physical systems. Furthermore, the framework presented in \cite{Mamaghani2026ISAC} combines Stackelberg game theory and DRL to jointly optimize security, sensing performance, and energy efficiency. These studies demonstrate that information related to uncertainty and user security behavior can be incorporated into network association and decision-making processes.

After establishing a realistic channel model, extracting reference behavior, identifying suspicious nodes, and determining user associations, the remaining challenge is to adaptively adjust communication resources in response to channel and security conditions. Exhaustively searching all possible combinations of beam angles and transmit-power levels rapidly becomes computationally expensive as the number of users and control variables increases. RL provides a suitable approach to address this challenge because a learning agent can learn the relationship among network states, actions, and resulting performance through successive interactions with the environment, enabling development of an adaptive decision-making policy \cite{Geranmayeh2025DQN}. This characteristic makes RL particularly suitable for dynamic wireless environments in which channel conditions, interference, and network security states vary over time. Deep RL has also been applied to resource allocation and optimization problems in communication networks with high traffic demands \cite{Vivekanand2024DRL}. In addition, RL-based frameworks for channel access control in IEEE 802.11 standards and mIoT networks have been investigated in \cite{Ali2021MassiveIoT}, while the role of AI and big data in advanced eMBB, mMTC, uRLLC, and integrated space-air-ground-sea network scenarios has been discussed in \cite{Wang2020Channels}. In such a framework, RL can incorporate both channel and security-state information into the joint beam and power selection process, enabling a trade-off among communication performance, energy consumption, and security rather than solely maximizing the aggregate full-band link-rate score.

Digital-twin residuals have previously been investigated for wireless anomaly
detection. Krause et al.\ compare DT-predicted received signal strengths with
distributed measurements to decide whether an anomaly, including a jammer, is
present \cite{Krause2023DTAnomaly}. Ruah et al.\ develop a Bayesian framework
for DT-based wireless monitoring and control that explicitly accounts for
model uncertainty \cite{Ruah2023BayesianDT}. Accordingly, the novelty claimed
here is not the generic comparison of expected and observed network
quantities. The present work addresses the narrower problem of ranking
candidate malicious receivers, under the assumptions that an attack is present
and that $N_{\mathrm{Att}}$ is known, while jointly adapting two transmit beams
and two transmit-power levels using BGQL. Its distinguishing components are
the multi-feature spatial-footprint score, temporal evidence aggregation, and
the use of that evidence in beam-power control.

To address this gap, this paper presents a security-oriented framework for adaptive resource management in an urban wireless network. The propagation environment is modeled using ray-tracing and realistic antenna patterns, while network behavior in the absence of attacks is used as an expected reference state against which the observed behavior under attacking is compared. Deviations in performance metrics are used to extract evidence associated with suspicious nodes, and this evidence is aggregated over time to update the system's belief regarding potential attackers. The resulting information is then incorporated into the decision-making process together with the channel state, while Q-learning and SARSA are employed for adaptive beam selection and transmit-power control. In each Monte Carlo run, the identity of the attacker or attackers is varied, while identical attack conditions are maintained across the compared methods to ensure a fair evaluation. 
Although the present study shares the broader research theme of secure wireless communications with the authors' previous studies \cite{Geranmayeh2026ISAC,Geranmayeh2026Bayesian}, it addresses a distinct problem and introduces a substantially different system model, methodology, and evaluation framework. Specifically, this work develops an SBR ray-tracing-based nominal-reference and attacked-network framework in which the received jamming power is modeled separately for each attacker-receiver link. It further proposes a multi-feature attacker-localization mechanism based on predicted jamming-footprint matching, temporal evidence aggregation, and belief updating, without providing the active attacker identity or the modified beam to the detector. In addition, the RL formulation jointly optimizes transmitter beams and transmit-power levels using a multi-objective reward that accounts for communication rate, power consumption, and belief confidence. The evaluation also considers six performance metrics across four controlled adversarial scenarios and includes paired statistical tests and effect-size analysis. These models, objectives, and experiments distinguish the present work from the two earlier studies, which are cited solely to clarify the relationship among the authors' studies in this broader research area.

The main contributions of this paper are summarized as follows:

\begin{itemize}
\item A physical model-based framework is presented to evaluate attack-free reference behavior and attacked simulated behavior. Unlike models that characterize the channel solely through statistical approximations, the legitimate transmitter-receiver propagation environment in this study is modeled using ray-tracing based on the Shooting and Bouncing Rays (SBR) method, together with reflection, diffraction, spatial shadowing, and angle-dependent antenna patterns \cite{Geok2018RayTracing,Ji2021RayTracingIndoor,Hsiao2017RayTracing}. This model is used to establish reference network behavior in the absence of attackers and to compare it with the observed network state under attack. In addition to changes in the attacker-receiver beam configuration, the received jamming power is calculated separately for each attacker-receiver link based on the corresponding channel gain. Therefore, the effect of an attacker is not represented by a fixed and uniform interference level across all receivers. This framework enables the spatial and channel-dependent effects of jamming on SINR and the aggregate full-band link-rate score to be investigated under the stated propagation models.

\item A multi-feature mechanism for attacker localization is developed based on temporal evidence aggregation and belief updating. Rather than detecting an attacker based on an instantaneous degradation in a single metric, the proposed framework compares attack-induced deviations in link rate and SINR, together with network disturbance characteristics, against predicted footprints of potential jamming sources. Importantly, neither the active attacker index nor the beam modified by the attacker is provided to the detector. The resulting evidence is accumulated over time using an exponential forgetting mechanism and is subsequently used to construct a belief distribution over suspicious nodes. The $N_{\mathrm{Att}}$ nodes with the highest belief values are then identified as attackers, where $N_{\mathrm{Att}}$ is supplied to every evaluated method. Unlike the Greedy detector employed as a baseline, which relies solely on instantaneous SINR degradation and does not incorporate fingerprint matching, temporal aggregation, or learned belief, the proposed method exploits cumulative evidence to achieve more stable attacker detection.

\item A multi-objective RL framework is designed for the joint optimization of beamforming and transmit power under adversarial conditions. BGQL and SARSA are implemented to learn adaptive beam-power policies. Unlike approaches that employ RL solely for beam selection or resource allocation \cite{Geranmayeh2025DQN,Vivekanand2024DRL}, the proposed state space incorporates information regarding suspicious nodes, detection confidence, channel quality, previous power states, and beam separation, while the action space includes joint beam selection and transmit-power levels for the two transmitters. Four transmit-power levels of $15$, $20$, $25$, and $30~\mathrm{dBm}$ are considered, which, together with $25$ angular options, form a joint beam-power decision space. Furthermore, a multi-objective reward function simultaneously incorporates communication rate, transmit-power cost, and belief confidence into the decision-making process. Consequently, the learned policy does not solely maximize the aggregate full-band link-rate score but instead balances communication performance, energy consumption, and security awareness.

\item A comprehensive and controlled evaluation is conducted to assess the performance under increasing network size and adversarial complexity of the proposed framework. BGQL and SARSA are evaluated under four network scenarios comprising $4$ receivers with $1$ attacker, $6$ receivers with $1$ attacker, $6$ receivers with $2$ attackers, and $8$ receivers with $2$ attackers. Attacker identities vary across independent Monte Carlo runs but remain fixed within each run, and the same attacker set is used for BGQL, SARSA, Greedy, and Random to ensure comparison under identical attack conditions. Performance is evaluated using six metrics: aggregate full-band link-rate score, attacker detection accuracy with a known number of attackers, total device power, aggregate radiated transmit power, rate-to-device-power score, and execution time. In addition to mean performance and variability, paired statistical tests and effect sizes are used to determine whether the observed differences among the evaluated algorithms are statistically significant.
\end{itemize}

\section{System Model and Architecture}
\label{sec:system_model}

This section presents the considered wireless network architecture and its propagation, antenna, association, energy, and adversarial models. The system represents an urban millimeter-wave (mmWave) network in which beam-dependent transmission and reception, transmitter-receiver association, transmit-power selection, and adversarial interference jointly determine the communication performance. A Digital-Twin (DT)-assisted reference representation is further employed to characterize the expected network behavior under intended operating conditions and to provide a baseline against which the attacked simulation response under adversarial conditions can be evaluated.

\subsection{Network Architecture and Urban Propagation Environment}
\label{subsec:network_architecture}
The considered network consists of $N_{\mathrm{TX}}=2$ transmitters
(TXs) and $N_{\mathrm{RX}}\in\{4,6,8\}$ receiver nodes (RXs). The system
operates at a carrier frequency of $f_c=28~\mathrm{GHz}$ over a bandwidth
of $B=400~\mathrm{MHz}$. Node locations and antenna heights are generated once at the
beginning of each complete simulation execution. Initial horizontal
coordinates are sampled within a $300\times300~\mathrm{m}^2$
Cartesian region, with $x,y\in[-150,150]~\mathrm{m}$.
The selected RX coordinates are subsequently perturbed by
Gaussian offsets with a standard deviation of $15~\mathrm{m}$
in each horizontal dimension.
TX heights are sampled from $8$-$23~\mathrm{m}$, and RX heights
from $1.2$-$2.2~\mathrm{m}$.
The four scenarios share the same two TXs and use the first
4, 6, or 8 RXs of the generated eight-RX deployment.
Within each scenario, node locations, antenna heights, and the
shadowing field remain fixed across all 30 Monte Carlo runs.
The coordinates in Table~\ref{tab:node_locations} illustrate
one generated deployment. The main network and propagation parameters are summarized
in Table~\ref{tab:system_parameters}.

To represent a realistic urban deployment, the Cartesian node
coordinates generated at the beginning of the simulation are mapped to geographical
coordinates using the reference point
$(51.5139^{\circ}\mathrm{N},\,7.4653^{\circ}\mathrm{E})$, corresponding
to downtown Dortmund, Germany. The resulting geographical locations
are aligned with the urban geometry obtained from OpenStreetMap (OSM),
which provides the site-specific environment used for propagation
modeling.

Four network scenarios are considered to investigate the effect of network size and adversarial activity. Scenario~1 contains four RXs and one attacker, Scenario~2 contains six RXs and one attacker, Scenario~3 contains six RXs and two attackers, and Scenario~4 contains eight RXs and two attackers. Independent Monte Carlo realizations are performed for each scenario. The attacker identities vary across Monte Carlo realizations but remain fixed throughout the iterations of an individual realization. Furthermore, the same attacker configuration is applied to all evaluated methods within each realization, providing a common basis for comparison.

The principal system and simulation parameters are summarized in
Table~\ref{tab:system_parameters}.

\renewcommand{\arraystretch}{1.15}
\begin{table*}[!t]
\centering\footnotesize
\caption{Main System and Simulation Parameters}
\label{tab:system_parameters}
\begin{tabular}{p{0.44\textwidth} p{0.17\textwidth} p{0.30\textwidth}}
\hline
\textbf{Parameter} & \textbf{Symbol} & \textbf{Value} \\
\hline

Number of transmitters
    & $N_{\mathrm{TX}}$ & $2$ \\

Number of receivers
    & $N_{\mathrm{RX}}$ & $4,6,8$ \\

Number of attackers
    & $N_{\mathrm{Att}}$ & $1,2$ \\

Network scenarios $(N_{\mathrm{RX}},N_{\mathrm{Att}})$
    & -- & $(4,1),(6,1),(6,2),(8,2)$ \\

Carrier frequency
    & $f_c$ & $28~\mathrm{GHz}$ \\

Bandwidth
    & $B$ & $400~\mathrm{MHz}$ \\

Number of antenna elements
    & $N_{\mathrm{ant}}$ & $10$ \\

Deployment area
    & -- & $300\times300~\mathrm{m}^2$ \\

TX height
    & $h_{\mathrm{TX}}$ & $8$--$23~\mathrm{m}$ \\

RX height
    & $h_{\mathrm{RX}}$ & $1.2$--$2.2~\mathrm{m}$ \\

Dortmund reference coordinate
    & -- & $51.5139^{\circ}\mathrm{N},\,7.4653^{\circ}\mathrm{E}$ \\

Ray-tracing method
    & -- & SBR \\

Maximum number of reflections
    & -- & $1$ \\

Maximum number of diffractions
    & -- & $1$ \\

Building/terrain material
    & -- & Concrete \\

Unfiltered Gaussian shadow-field standard deviation
    & $\sigma_{\mathrm{shadow}}$ & $7~\mathrm{dB}$ \\

Shadowing correlation distance
    & $d_{\mathrm{corr}}$ & $50~\mathrm{m}$ \\

Beam-angle range
    & $\Theta$ & $[-60^{\circ},60^{\circ}]$ \\

Angular resolution
    & $\Delta\theta$ & $5^{\circ}$ \\

Number of candidate beam directions
    & $N_{\theta}$ & $25$ \\

TX power levels
    & $\mathcal{P}$ & $\{15,20,25,30\}~\mathrm{dBm}$ \\

Attacker transmit power
    & $P_J$ & $20~\mathrm{dBm}$ \\

Thermal noise density
    & $N_0$ & $-174~\mathrm{dBm/Hz}$ \\

RX noise figure
    & $\mathrm{NF}_{\mathrm{UE}}$ & $13~\mathrm{dB}$ \\

PA efficiency
    & $\eta_{\mathrm{PA}}$ & $0.38$ \\

BS static power
    & $P_{\mathrm{static}}^{\mathrm{BS}}$ & $2~\mathrm{W}$ \\

RF-chain power
    & $P_{\mathrm{RF}}$ & $0.15~\mathrm{W}$ \\
    
Number of RF chains per TX
    & $N_{\mathrm{RF}}$ & $1$ \\
    
Baseband-processing power
    & $P_{\mathrm{BB}}$ & $0.30~\mathrm{W}$ \\

RX static power
    & $P_{\mathrm{static}}^{\mathrm{RX}}$ & $0.10~\mathrm{W}$ \\

RX RF-chain power
    & $P_{\mathrm{RF}}^{\mathrm{RX}}$ & $0.05~\mathrm{W}$ \\
\hline
\end{tabular}
\end{table*}

The corresponding values, including the beam range, power levels, attacker power, noise parameters, and hardware-power parameters, reflect the implemented system configuration and are summarized in Table~\ref{tab:system_parameters}.

The mmWave propagation environment is modeled using site-specific ray-tracing based on the Shooting and Bouncing Rays (SBR) method. The urban geometry and surrounding obstacles are explicitly incorporated, with concrete used for the building and terrain materials. For each TX-RX link, the ray-tracing procedure considers up to one reflection and one diffraction and provides the path loss, propagation phase, angle of departure (AoD), and angle of arrival (AoA) of the resulting propagation components. Therefore, whenever valid SBR paths are available, the propagation information used in the subsequent channel, SINR, and link-rate calculations is obtained directly from the site-specific ray-tracing model.

For numerical robustness, a line-of-sight free-space fallback is employed whenever the ray-tracing procedure does not return a valid propagation path for a particular TX-RX pair. Let $d_{t,r}$ denote the three-dimensional distance between TX $t$ and RX $r$. The corresponding free-space path loss is

\begin{equation}
L_{t,r}^{\mathrm{FS}}
=
20\log_{10}
\left(
\frac{4\pi d_{t,r}}{\lambda}
\right),
\label{eq:fspl}
\end{equation}

where

\begin{equation}
\lambda = \frac{c}{f_c}
\label{eq:wavelength}
\end{equation}

is the carrier wavelength. The corresponding free-space propagation phase is
determined as

\begin{equation}
\phi_{t,r}^{\mathrm{FS}}
=
-\frac{2\pi d_{t,r}}{\lambda}.
\label{eq:fs_phase}
\end{equation}

The associated AoD and AoA are obtained geometrically from the relative TX
and RX coordinates. This fallback is a numerical modeling assumption and is
not evidence that an unobstructed LOS path exists when the SBR solver returns
no path. The reported conclusions therefore inherit any bias introduced by
fallback links.

In addition to the deterministic site-specific propagation components, spatially correlated shadow fading is superimposed on the channel model. The initial i.i.d. Gaussian field has a standard deviation of
$\sigma_{\mathrm{shadow}}=7~\mathrm{dB}$ and is spatially filtered using an exponential kernel with decay-distance parameter $d_{\mathrm{corr}}=50~\mathrm{m}$.
Filtering changes the field variance, and no subsequent
variance renormalization is applied. The correlated field is obtained by spatially filtering an i.i.d.\ Gaussian field using an exponentially decaying kernel,

\begin{equation}
K(\Delta x)
=
\exp
\left(
-\frac{|\Delta x|d_x}{d_{\mathrm{corr}}}
\right),
\label{eq:shadow_kernel}
\end{equation}

where $d_x$ denotes the spatial grid resolution. For a link between TX $i$ and RX $j$, the corresponding link-level shadowing contribution is constructed from the spatial field values associated with the two endpoints as

\begin{equation}
S_{i,j}
=
S_{\mathrm{TX},i}
+
S_{\mathrm{RX},j}.
\label{eq:link_shadowing}
\end{equation}

This formulation introduces spatial correlation into the shadowing process so that geographically close nodes experience related, rather than statistically independent, large-scale channel variations. The shadowing parameters and their role in the propagation model are part of the original system formulation.

\subsection{Directional Beamforming, Channel Model, and RX Association}
\label{subsec:beamforming_channel}

Beam-dependent transmit and receive antenna responses are incorporated through precomputed directional antenna radiation patterns. The antenna responses are stored prior to the network simulation and subsequently selected according to the TX and RX beam indices. Consequently, the online optimization does not redesign the antenna structure or regenerate a radiation pattern at every iteration; instead, a beamforming decision corresponds to selecting one of the available precomputed directional responses. This representation is consistent with the beam-index-based channel formulation used throughout the simulation.

A discrete beam codebook containing 25 steering angles from
$-60^{\circ}$ to $60^{\circ}$, in steps of $5^{\circ}$,
is considered:

\begin{equation}
\Theta
=
\{
-60^{\circ},
-55^{\circ},
\ldots,
0^{\circ},
\ldots,
55^{\circ},
60^{\circ}
\},
\label{eq:beam_codebook}
\end{equation}

which results in

\begin{equation}
N_{\theta}=25
\label{eq:number_beams}
\end{equation}

candidate beam directions. The corresponding directional antenna responses are precomputed and stored prior to the network simulation. During online operation, the selected beam index identifies the appropriate TX or RX radiation pattern.
The selected TX and RX antenna patterns are evaluated at the ray-specific angles of departure (AoDs) and angles of arrival (AoAs), respectively, and their corresponding gains are incorporated into the effective channel calculation. The beam range and resolution correspond to the implemented codebook.

For each available propagation component $k$, the beam-dependent TX and RX gains are extracted from the selected antenna patterns at the corresponding AoD and AoA. The complex channel contribution of path $k$ between TX $t$ and RX $r$ is represented as

\begin{align}
    h_{t,r}^{(k)} = 10^{\frac{G_t^{(k)}+G_r^{(k)}-L_{t,r}^{(k)}-S_{t,r}}{20}}e^{j\phi_{t,r}^{(k)}}, \label{eq:path_channel}
\end{align}

where $G_t^{(k)}$ and $G_r^{(k)}$ are the beam-dependent transmit and receive antenna gains, respectively; $L_{t,r}^{(k)}$ denotes the path loss of propagation component $k$; $S_{t,r}$ represents the spatial shadowing contribution; and $\phi_{t,r}^{(k)}$ is the corresponding propagation phase.

The available multipath components are combined coherently, and the effective channel power gain used in the subsequent received-power and SINR calculations is therefore

\begin{align}
    H_{t,r} = \left| \sum_k h_{t,r}^{(k)} \right|^2. \label{eq:channel_power_gain}
\end{align}

Thus, $H_{t,r}$ jointly incorporates the selected TX and RX beam responses, site-specific propagation loss, propagation phase, coherent multipath combination, and spatially correlated shadow fading.

Each RX is subsequently associated with one of the two TXs. Rather than assigning an RX solely according to the strongest individual link, TX-RX association is formulated as a non-cooperative game with stochastic best-response dynamics. For RX $j$, the utility associated with selecting TX $i$ is

\begin{equation}
U_j(i)
=
\log_2
\left(
1+\mathrm{SINR}_j(i)
\right)
-
\beta\log(N_i+1),
\label{eq:association_utility}
\end{equation}

where $N_i$ denotes the number of RXs that would be associated with TX $i$ after receiver $j$ selects it, including receiver $j$, while $\beta$ controls the load-balancing penalty. The first term favors associations providing higher communication quality, whereas the second term discourages excessive concentration of RXs at the same transmitter.
The association decision is updated according to a Softmax stochastic best response,

\begin{equation}
p_j(i)
=
\frac{
\exp\left(U_j(i)/\tau\right)
}{
\displaystyle
\sum_{m=1}^{N_{\mathrm{TX}}}
\exp\left(U_j(m)/\tau\right)
},
\label{eq:softmax_association}
\end{equation}

where $\tau$ determines the stochasticity of the response. In the implemented
association procedure,

\begin{equation}
\beta=0.8,
\qquad
\tau=0.7,
\label{eq:association_parameters}
\end{equation}

and a maximum of 15 game iterations is employed. Accordingly, TX-RX association is determined during network initialization through the stochastic association game rather than through a simple strongest-link rule.

For each RX, the desired received signal, interference from the non-serving transmitter, and adversarial jamming contribution are evaluated separately. Let $s_r$ denote the transmitter serving RX $r$. The desired received signal power is

\begin{equation}
P_r^{\mathrm{sig}}
=
P_{s_r}H_{s_r,r},
\label{eq:signal_power}
\end{equation}

whereas the aggregate interference generated by the non-serving transmitters is

\begin{equation}
P_r^{\mathrm{int}}
=
\sum_{\substack{t=1\\t\neq s_r}}^{N_{\mathrm{TX}}}
P_t H_{t,r}.
\label{eq:interference_power}
\end{equation}

The received SINR is consequently

\begin{equation}
\gamma_r
=
\frac{
P_r^{\mathrm{sig}}
}{
\sigma^2
+
P_r^{\mathrm{int}}
+
J_r
}
=
\frac{
P_{s_r}H_{s_r,r}
}{
\sigma^2
+
\displaystyle\sum_{t\neq s_r}P_tH_{t,r}
+
J_r
}.
\label{eq:sinr}
\end{equation}

Here, $J_r$ denotes the aggregate adversarial jamming power received at RX $r$. Under the reference non-adversarial condition, $J_r=0$.

The thermal noise density is set to $N_0=-174~\mathrm{dBm/Hz}$, while the RX noise figure is $\mathrm{NF}_{\mathrm{UE}}=13~\mathrm{dB}$. The corresponding
receiver noise power is obtained as

\begin{equation}
\sigma_{\mathrm{dBm}}^2
=
N_0
+
10\log_{10}(B)
+
\mathrm{NF}_{\mathrm{UE}},
\label{eq:noise_power}
\end{equation}

and is converted to linear units before being included in
\eqref{eq:sinr}.

For each RX $r$, the standalone full-band achievable-link-rate metric is
calculated as
\begin{equation}
C_r
=
B\log_2(1+\gamma_r).
\label{eq:receiver_capacity}
\end{equation}
The corresponding aggregate full-band link-rate score is defined as
\begin{equation}
C_{\mathrm{tot}}
=
\sum_{r=1}^{N_{\mathrm{RX}}} C_r.
\label{eq:total_capacity}
\end{equation}
This score depends jointly on the beam-dependent channel gains, transmit
powers, association decisions, inter-transmitter interference, thermal
noise, and adversarial interference. Because the considered model does
not explicitly specify orthogonal bandwidth allocation, simultaneous
independent spatial streams, or an intra-transmitter multiuser signal
model, $C_{\mathrm{tot}}$ is used as an optimization and comparative
performance score. It is therefore not interpreted as the simultaneous
Shannon sum capacity of the physical network.

\subsection{Power Consumption and Rate-to-Device-Power Model}
\label{subsec:power_model}

In addition to maximizing communication performance, the considered system explicitly accounts for the energy cost associated with the selected transmit powers. Each transmitter can operate at one of four discrete radiated power levels,

\begin{equation}
\mathcal{P}
=
\{15,20,25,30\}~\mathrm{dBm}.
\label{eq:power_levels}
\end{equation}

The device-power model incorporates both radiated and circuit-level power consumption. For each BS, the static circuit power is
$P_{\mathrm{static}}^{\mathrm{BS}}=2~\mathrm{W}$, the RF-chain consumption is $P_{\mathrm{RF}}=0.15~\mathrm{W}$, and the baseband-processing consumption is $P_{\mathrm{BB}}=0.30~\mathrm{W}$. The power-amplifier drain efficiency is $\eta_{\mathrm{PA}}=0.38$. On the receiver side, each RX consumes $P_{\mathrm{static}}^{\mathrm{RX}}=0.10~\mathrm{W}$ of static power and
$P_{\mathrm{RF}}^{\mathrm{RX}}=0.05~\mathrm{W}$ for its RF chain. These values correspond to the implemented energy-consumption model.

For transmitter $t$, the BS device power is modeled as

\begin{equation}
P_{\mathrm{BS},t}
=
P_{\mathrm{static}}^{\mathrm{BS}}
+
N_{\mathrm{RF}}P_{\mathrm{RF}}
+
P_{\mathrm{BB}}
+
\frac{P_t}{\eta_{\mathrm{PA}}},
\label{eq:bs_power}
\end{equation}

The implemented model uses one RF chain per transmitter,
$N_{\mathrm{RF}}=1$. The radiated power $P_t$ is converted from dBm to watts
before the PA term is evaluated.

The aggregate receiver-side power consumption is

\begin{equation}
P_{\mathrm{RX}}
=
N_{\mathrm{RX}}
\left(
P_{\mathrm{static}}^{\mathrm{RX}}
+
P_{\mathrm{RF}}^{\mathrm{RX}}
\right),
\label{eq:rx_power}
\end{equation}

and the total network device power is therefore

\begin{equation}
P_{\mathrm{dev}}
=
\sum_{t=1}^{N_{\mathrm{TX}}}
P_{\mathrm{BS},t}
+
N_{\mathrm{RX}}
\left(
P_{\mathrm{static}}^{\mathrm{RX}}
+
P_{\mathrm{RF}}^{\mathrm{RX}}
\right).
\label{eq:device_power}
\end{equation}

The rate-to-device-power score is defined as
\begin{equation}
\eta_{\mathrm{RP}}
=
\frac{C_{\mathrm{tot}}}{P_{\mathrm{dev}}}
\quad [\mathrm{bit/J}].
\label{eq:energy_efficiency}
\end{equation}
This score incorporates the considered BS and RX hardware-power terms. Because its numerator is the aggregate full-band link-rate score defined above, rather than delivered multiuser throughput, $\eta_{\mathrm{RP}}$ is used as a comparative rate-to-device-power score. It is not interpreted as the deployment-level energy efficiency of the physical network.

\subsection{Attacker and Jamming Model}
\label{subsec:attacker_model}

To evaluate the network under adversarial conditions, a subset of the RX nodes is assumed to behave maliciously. Depending on the considered network scenario, either one or two nodes are designated as attackers. The attacker identities vary across independent Monte Carlo realizations but remain unchanged during the iterations belonging to a particular realization. Each active attacker transmits a jamming signal with a fixed power of

\begin{equation}
P_J = 20~\mathrm{dBm}.
\label{eq:Attacker_power}
\end{equation}

The adversarial effect is spatially dependent; consequently, the received jamming power is not assumed to be identical across RXs. Instead, an attacker-to-receiver channel-gain matrix is constructed to characterize the disturbance footprint generated by each potential attacker. The baseline formulation models a separate attacker gain and aggregate $J_r$ for each receiver.

For attacker $a$ and RX $r$, $a\neq r$, the three-dimensional separation is calculated as

\begin{equation}
d_{a,r}
=
\sqrt{
(x_r-x_a)^2
+
(y_r-y_a)^2
+
(z_r-z_a)^2
}.
\label{eq:Attacker_distance}
\end{equation}

The corresponding free-space path loss of the attacker link is

\begin{equation}
L_{a,r}^{J}
=
20\log_{10}
\left(
\frac{4\pi d_{a,r}}{\lambda}
\right).
\label{eq:Attacker_fspl}
\end{equation}

In accordance with the implemented attacker-channel representation, the shadowing contribution of an attacker-receiver pair is approximated from the spatial shadowing values associated with its two endpoints:

\begin{equation}
S_{a,r}^{J}
=
\frac{1}{2}
\left(
S_{\mathrm{RX},a}
+
S_{\mathrm{RX},r}
\right).
\label{eq:Attacker_shadowing}
\end{equation}

The total attacker-link attenuation is constrained to remain non-negative,

\begin{equation}
A_{a,r}^{J}
=
\max
\left(
L_{a,r}^{J}
+
S_{a,r}^{J},
0
\right),
\label{eq:Attacker_attenuation}
\end{equation}

and the corresponding linear attacker-link power gain becomes

\begin{equation}
G_{a,r}^{J}
=
10^{-A_{a,r}^{J}/10}
=
10^{
-\frac{
\max\left(
L_{a,r}^{J}
+
S_{a,r}^{J},
0
\right)
}{10}
}.
\label{eq:Attacker_gain}
\end{equation}

The diagonal terms of the attacker-gain matrix satisfy

\begin{equation}
G_{a,a}^{J}=0,
\label{eq:Attacker_diagonal}
\end{equation}

since a malicious RX is not modeled as generating a jamming contribution toward itself.

For the set of active attackers $\mathcal{A}_{\mathrm{att}}$, the aggregate jamming power observed by RX $r$ is consequently

\begin{equation}
J_r
=
P_J
\sum_{\substack{
a\in\mathcal{A}_{\mathrm{att}}\\
a\neq r
}}
G_{a,r}^{J},
\label{eq:aggregate_jamming}
\end{equation}

where $P_J$ is converted from dBm to watts before the received-power calculation. Thus, different RXs experience different levels of adversarial interference depending on their spatial relationship with the active attackers.

The adversarial behavior is not limited to the additional jamming contribution. A malicious RX also deviates from its intended receive-beam configuration. Consequently, the attack affects the simulated network model through two coupled mechanisms: an abnormal RX beam configuration at the malicious node and spatially distributed jamming interference affecting the surrounding receivers. The active attacker set is used internally only to generate the simulated attack and to calculate the post-learning localization-recall metric. It is not supplied to the anomaly detector, the RL state, the action-selection mechanism, or the reward function.
The attacker link is a simplified equivalent-isotropic channel consisting of free-space loss and endpoint-based shadowing. It does not use the site-specific SBR paths or the directional antenna patterns employed for the legitimate links. Jammer self-interference is set to zero through $G_{a,a}^{J}=0$. The jammer’s transmit power is also excluded from $P_{\mathrm{dev}}$ because it represents uncontrolled adversarial energy. These assumptions define the evaluated threat model and should not be interpreted as a general or fully hardware-realistic jammer model.

\subsection{DT-Assisted Reference Network Representation}
\label{subsec:dt_reference}

The term DT reference is used here in a restricted, simulation-based sense.
For each beam-power action, the same network simulator is evaluated twice:
once with the intended RX beams and no jammer, and once with the simulated
adversarial modifications. The first evaluation supplies the attack-free
reference and the second supplies the attacked simulated response. No
physical-network telemetry, online virtual-physical synchronization,
measurement-calibration procedure, or DT-model mismatch is included.
Consequently, the evaluation represents a matched-model simulation proof of
concept rather than a deployed physical-network DT.

More specifically, for each beam and transmit-power decision, the same simulated network model is evaluated twice. First, the DT-assisted reference evaluates the intended network configuration without adversarial modification. The intended RX beam configuration is retained and the jammer contribution is disabled, i.e.,

\begin{equation}
J_r=0.
\label{eq:no_jamming_dt}
\end{equation}

This first evaluation provides the expected receiver-level link rate, SINR, desired received signal power, and interference power. The expected reference state is therefore represented as

\begin{equation}
\mathcal{S}_{\mathrm{exp}}
=
\left\{
\mathbf{C}_{\mathrm{exp}},
\boldsymbol{\gamma}_{\mathrm{exp}},
\mathbf{P}_{\mathrm{sig,exp}},
\mathbf{P}_{\mathrm{int,exp}}
\right\}.
\label{eq:expected_state}
\end{equation}

Subsequently, the simulated attack is introduced while preserving the same network topology, propagation realization, TX-RX association, TX beam decision, and transmit-power configuration. For each malicious node, its receive beam is changed from the intended configuration, and the corresponding spatial jamming contribution $J_r$ is generated according to the attacker-to-receiver gain matrix defined in
\eqref{eq:Attacker_gain}-\eqref{eq:aggregate_jamming}. The simulated network model is then evaluated for a second time, resulting in the attacked simulation state

\begin{equation}
\mathcal{S}_{\mathrm{act}}
=
\left\{
\mathbf{C}_{\mathrm{act}},
\boldsymbol{\gamma}_{\mathrm{act}},
\mathbf{P}_{\mathrm{sig,act}},
\mathbf{P}_{\mathrm{int,act}},
\mathbf{P}_{\mathrm{jam,act}}
\right\}.
\label{eq:actual_state}
\end{equation}

The distinction between the two evaluations is therefore not caused by a change in the underlying propagation environment or topology. Instead, the expected and attacked simulation states differ primarily through the adversarial RX-beam deviation and the additional spatial jamming contribution. This controlled
comparison makes it possible to associate deviations in network behavior with potential malicious activity.

The DT-assisted reference-versus-attacked comparison can consequently be expressed as

\begin{equation}
\mathcal{S}_{\mathrm{exp}}
\longleftrightarrow
\mathcal{S}_{\mathrm{act}}.
\label{eq:dt_comparison}
\end{equation}

The link-rate and SINR residuals are evaluated together with the desired-signal and interference-power residuals and the spatial disturbance patterns associated with the candidate attacker locations. This reference-versus-attacked comparison provides the basis for the subsequent detection stage.

The resulting deviations are subsequently supplied to the attacker-detection and temporal belief-updating procedure. The resulting belief estimates are subsequently incorporated into the adaptive framework for joint beam and transmit-power optimization. These operations constitute the proposed decision-making methodology rather than the underlying physical system model. Therefore, the multi-feature anomaly formulation, spatial attacker evidence, temporal belief update, RL state and action spaces, reward formulation, and the Q-learning and SARSA optimization procedures are introduced separately in
Section~\ref{sec:methodology}.
\section{Proposed Methodology}
\label{sec:methodology}

The proposed method combines DT-based network monitoring with  attacker localization and reinforcement learning (RL) for joint beam and  transmit-power control. The DT is used to establish the network response  expected under normal operating conditions. This reference is compared with the attacked response of the same simulated network model, and the resulting deviations are used to estimate which receivers are likely to be malicious. The estimated attacker information is then incorporated into the RL state and, in the Q-learning scheme, is also used to guide exploratory beam selection.

\subsection{DT-Assisted Network Monitoring}
\label{subsec:dt_monitoring}

For a given beam and transmit-power configuration, the DT provides the
expected network response in the absence of adversarial interference.
The relevant receiver-level quantities are represented by

\begin{equation}
\mathcal{X}^{\mathrm{exp}}
=
\left\{
\mathbf{C}^{\mathrm{exp}},
\boldsymbol{\gamma}^{\mathrm{exp}},
\mathbf{P}_{\mathrm{sig}}^{\mathrm{exp}},
\mathbf{P}_{\mathrm{int}}^{\mathrm{exp}}
\right\},
\label{eq:expected_network_quantities}
\end{equation}

where $\mathbf{C}^{\mathrm{exp}}$ and
$\boldsymbol{\gamma}^{\mathrm{exp}}$ denote the expected receiver link rates and SINRs, respectively, while
$\mathbf{P}_{\mathrm{sig}}^{\mathrm{exp}}$ and
$\mathbf{P}_{\mathrm{int}}^{\mathrm{exp}}$ contain the corresponding
desired-signal and inter-transmitter interference powers.

Under attack, malicious receivers may use receive beams different from
their intended configurations while also generating additional jamming
interference. The detector is not assumed to separately observe the
inter-transmitter interference and the jamming power. Instead, it uses
their aggregate disturbance, defined as

\begin{equation}
\mathbf{P}_{\mathrm{dist}}^{\mathrm{act}}
=
\mathbf{P}_{\mathrm{int}}^{\mathrm{act}}
+
\mathbf{P}_{\mathrm{jam}}^{\mathrm{act}}.
\label{eq:actual_aggregate_disturbance}
\end{equation}

The two components on the right-hand side of
Eq.~\eqref{eq:actual_aggregate_disturbance} are separately available
only inside the simulator for physical channel calculation. In the
reported experiments, the detector receives
$\mathbf{P}_{\mathrm{dist}}^{\mathrm{act}}$ and the remaining quantities in
$\mathcal{X}^{\mathrm{act}}$ directly from the simulator without measurement
or estimation error. Consequently, the results quantify matched-model
simulation performance and do not establish robustness to imperfect receiver
estimation or DT calibration.

Accordingly, the observed receiver-level quantities under attack are
represented by

\begin{equation}
\mathcal{X}^{\mathrm{act}}
=
\left\{
\mathbf{C}^{\mathrm{act}},
\boldsymbol{\gamma}^{\mathrm{act}},
\mathbf{P}_{\mathrm{sig}}^{\mathrm{act}},
\mathbf{P}_{\mathrm{dist}}^{\mathrm{act}}
\right\}.
\label{eq:actual_network_quantities}
\end{equation}

The detector operates on the deviations between
$\mathcal{X}^{\mathrm{exp}}$ and $\mathcal{X}^{\mathrm{act}}$. It does
not receive the active attacker indices, the malicious receive-beam
choices, or the separately calculated jamming-power component.

The main processing sequence is

\begin{equation}
\begin{split}
\text{DT reference}
&\rightarrow
\text{network observation}
\\
&\rightarrow\text{anomaly extraction}
\\
&\rightarrow
\text{temporal evidence}
\rightarrow
\text{attacker belief}
\\
&\rightarrow
\text{RL decision}.
\end{split}
\label{eq:framework_flow}
\end{equation}

This structure allows the resource-allocation decisions to reflect both
the current communication conditions and the accumulated evidence of
malicious behavior.

\subsection{Multi-Feature Attacker Localization}
\label{subsec:attacker_localization}

An attacker cannot be reliably identified based on a change in only one
performance metric. Variations in propagation and inter-transmitter
interference can also produce temporary reductions in link rate or
SINR. Accordingly, attacker localization is based on a combination of
local performance deviations and the spatial pattern of the excess
disturbance observed across the receivers.

For receiver $r$, the normalized link-rate degradation is

\begin{equation}
D_r^{C}
=
\left[
\frac{
\max\left(0,C_r^{\mathrm{exp}}-C_r^{\mathrm{act}}\right)
}{
\left|C_r^{\mathrm{exp}}\right|+\varepsilon
}
\right]_{0}^{1},
\label{eq:capacity_anomaly}
\end{equation}

where $[\cdot]_{0}^{1}$ denotes clipping to the interval $[0,1]$. The
corresponding normalized SINR degradation is

\begin{equation}
D_r^{\gamma}
=
\left[
\frac{
\max\left(0,\gamma_r^{\mathrm{exp}}-\gamma_r^{\mathrm{act}}\right)
}{
\left|\gamma_r^{\mathrm{exp}}\right|+\varepsilon
}
\right]_{0}^{1}.
\label{eq:sinr_anomaly}
\end{equation}

Throughout the manuscript, $\varepsilon$ denotes a
context-dependent numerical safeguard, rather than a single
physical constant. The implementation adds $10^{-12}$~bit/s
to the rate-drop denominator and a dimensionless $10^{-12}$
to the SINR-drop denominator. Power denominators in the SINR
and rate-to-device-power calculations use $10^{-15}$~W.
Received signal powers are floored at $10^{-20}$~W before
computing their logarithmic ratio. Spatial fingerprint-matching denominators use $10^{-30}$ in their respective native units. Power-cost and belief-reward normalizations use MATLAB's machine epsilon, $\mathrm{eps}\approx2.22\times10^{-16}$,
in the corresponding native units. These safeguards are
distinct from the exploration probability $\epsilon_t$.
The desired-signal power mismatch is measured in the logarithmic domain
as

\begin{equation}
\Delta P_r^{\mathrm{dB}}
=
\left|
10\log_{10}
\left(
\frac{
\max(P_{r,\mathrm{sig}}^{\mathrm{act}},10^{-20}\,\mathrm{W})
}{
\max(P_{r,\mathrm{sig}}^{\mathrm{exp}},10^{-20}\,\mathrm{W})
}
\right)
\right|,
\label{eq:signal_mismatch_db}
\end{equation}

and normalized according to

\begin{equation}
D_r^{P}
=
\left[
\frac{\Delta P_r^{\mathrm{dB}}}{10}
\right]_{0}^{1}.
\label{eq:power_anomaly}
\end{equation}

Unlike $D_r^{C}$ and $D_r^{\gamma}$, which describe degradation,
$D_r^{P}$ measures the magnitude of the desired-signal mismatch and can
therefore capture deviations in either direction.

Using the aggregate observed disturbance defined in
Eq.~\eqref{eq:actual_aggregate_disturbance}, the excess disturbance
relative to the DT reference is calculated as

\begin{equation}
\Delta\mathbf{I}
=
\left[
\mathbf{P}_{\mathrm{dist}}^{\mathrm{act}}
-
\mathbf{P}_{\mathrm{int}}^{\mathrm{exp}}
\right]_{+},
\label{eq:excess_disturbance}
\end{equation}

where $[\cdot]_{+}$ retains the non-negative components.

For a candidate attacker $a$, the expected spatial jamming footprint is
given by

\begin{equation}
\mathbf{g}_{a}
=
\left[
G_{a,1}^{J},
\ldots,
G_{a,N_{\mathrm{RX}}}^{J}
\right],
\label{eq:Attacker_fingerprint}
\end{equation}

with $G_{a,a}^{J}=0$, since a node is not modeled as generating a
jamming contribution toward itself.

The agreement between the observed excess disturbance and the
candidate footprint is first quantified using the cosine similarity

\begin{equation}
\chi_a
=
\left[
\frac{
\Delta\mathbf{I}^{T}\mathbf{g}_{a}
}{
\|\Delta\mathbf{I}\|_2
\|\mathbf{g}_{a}\|_2+\varepsilon
}
\right]_{0}^{1}.
\label{eq:cosine_similarity}
\end{equation}

A non-negative scaling factor is then estimated as

\begin{equation}
\hat{\alpha}_{a}
=
\max
\left(
\frac{
\Delta\mathbf{I}^{T}\mathbf{g}_{a}
}{
\mathbf{g}_{a}^{T}\mathbf{g}_{a}+\varepsilon
},
0
\right),
\label{eq:scale_factor}
\end{equation}

and the quality of the resulting footprint fit is measured by

\begin{equation}
F_a
=
\left[
1-
\frac{
\left\|
\Delta\mathbf{I}
-
\hat{\alpha}_{a}\mathbf{g}_{a}
\right\|_2
}{
\|\Delta\mathbf{I}\|_2+\varepsilon
}
\right]_{0}^{1}.
\label{eq:fit_quality}
\end{equation}

The spatial source score is defined as

\begin{equation}
S_a^{\mathrm{src}}
=
0.65\chi_a
+
0.35F_a.
\label{eq:source_score}
\end{equation}

When several jammers are active, the observed excess-disturbance vector is a
superposition of several footprints, whereas $S_a^{\mathrm{src}}$ fits one
candidate footprint at a time. The resulting score is therefore a heuristic
candidate ranking rather than an exact multi-source decomposition.
Finally, the instantaneous anomaly score assigned to candidate $a$ is

\begin{equation}
S_a
=
\left[
0.75S_a^{\mathrm{src}}
+
0.15D_a^{P}
+
0.05D_a^{\gamma}
+
0.05D_a^{C}
\right]_{0}^{1}.
\label{eq:multifeature_score}
\end{equation}

The spatial disturbance footprint therefore carries most of the localization weight. Link-rate and SINR residuals provide correlated engineered features because $C_r=B\log_2(1+\gamma_r)$ makes the capacity residual a
deterministic function of the corresponding SINR pair. They are retained with separate nonlinear normalizations, but they are not interpreted as independent evidence.

\subsection{Temporal Evidence and Attacker Belief}
\label{subsec:belief_update}

Instantaneous anomaly scores can vary considerably from one iteration
to the next. To reduce sensitivity to short-term fluctuations, the
score associated with each candidate is accumulated according to

\begin{equation}
E_a(t)
=
\rho E_a(t-1)
+
(1-\rho)S_a(t),
\label{eq:evidence_update}
\end{equation}

with

\begin{equation}
\rho=0.95.
\label{eq:evidence_forgetting}
\end{equation}

After an initial observation period of 200 iterations, the accumulated
evidence is centered around its receiver-level mean and scaled
according to

\begin{equation}
z_a(t)
=
\kappa
\left[
E_a(t)-\bar{E}(t)
\right],
\qquad
\kappa=12,
\label{eq:evidence_scaling}
\end{equation}

where

\begin{equation}
\bar{E}(t)
=
\frac{1}{N_{\mathrm{RX}}}
\sum_{j=1}^{N_{\mathrm{RX}}}
E_j(t).
\label{eq:mean_evidence}
\end{equation}

A normalized belief is then obtained using the Softmax mapping

\begin{equation}
b_a(t)
=
\frac{
\exp\left(z_a(t)\right)
}{
\displaystyle
\sum_{j=1}^{N_{\mathrm{RX}}}
\exp\left(z_j(t)\right)
}.
\label{eq:attacker_belief}
\end{equation}

For numerical stability, the maximum entry of $\mathbf{z}(t)$ is
subtracted from all its entries before evaluating the exponentials.
This operation does not change the resulting Softmax values.

The belief vector is written as

\begin{equation}
\mathbf{b}(t)
=
\left[
b_1(t),
\ldots,
b_{N_{\mathrm{RX}}}(t)
\right].
\label{eq:belief_vector}
\end{equation}

Here, $\mathbf{b}(t)$ represents a normalized operational measure of the
accumulated evidence associated with each candidate receiver. It is not
interpreted as an analytical Bayesian posterior.

At the end of run $m$, the method selects exactly
$N_{\mathrm{Att}}$ candidates. Its attacker detection accuracy, defined here as the fraction of the
$N_{\mathrm{Att}}$ true attackers included among the
$N_{\mathrm{Att}}$ selected candidates, is
\begin{equation}
\mathrm{Acc}_{m}
=
\frac{
\left|
\widehat{\mathcal{A}}_m
\cap
\mathcal{A}_{\mathrm{att},m}
\right|
}{
N_{\mathrm{Att}}
}.
\label{eq:detection_accuracy}
\end{equation}
The reported mean is
\begin{equation}
\overline{\mathrm{Acc}}
=
\frac{1}{M}
\sum_{m=1}^{M}
\mathrm{Acc}_{m}.
\label{eq:mean_detection_accuracy}
\end{equation}
All evaluated runs contain an attack, and the exact value of
$N_{\mathrm{Att}}$ is supplied to every method. This known-cardinality metric evaluates attacker identification; it is not conventional all-node classification accuracy and does not evaluate attack-presence detection, false-alarm probability, or unknown-count estimation. Uniform random selection has expected accuracy
$N_{\mathrm{Att}}/N_{\mathrm{RX}}$, equal to $25\%$, $16.7\%$, $33.3\%$, and
$25\%$ in the four scenarios, respectively.

\subsection{RL State and Joint Beam-Power Action Space}
\label{subsec:rl_state_action}

The RL state combines the estimated security condition with information
about channel quality and the recent resource configuration. At
iteration $t$, the state is represented by

\begin{equation}
s_t
=
\left(
\mathcal{B}_t,
q_t^{\mathrm{conf}},
q_t^{\mathrm{ch}},
q_t^{P},
q_t^{\Delta\theta}
\right),
\label{eq:rl_state}
\end{equation}

where $\mathcal{B}_t$ contains the $N_{\mathrm{Att}}$ receivers with
the largest current belief values. The remaining components represent
the belief confidence, channel quality, previous transmit-power
configuration, and separation between the two transmit beams,
respectively.

The number of possible suspected-attacker sets is

\begin{equation}
N_{\mathrm{sus}}
=
\binom{N_{\mathrm{RX}}}{N_{\mathrm{Att}}}.
\label{eq:suspected_sets}
\end{equation}

The confidence associated with the current suspected-attacker set is
defined as

\begin{equation}
B_t^{\mathrm{top}}
=
\sum_{a\in\mathcal{B}_t}
b_a(t),
\label{eq:belief_confidence}
\end{equation}

The top-set belief mass is directly quantized into five equal-width
confidence levels as
\begin{equation}
q_t^{\mathrm{conf}}
=
1+\min\!\left(4,\left\lfloor 5B_t^{\mathrm{top}}\right\rfloor\right).
\label{eq:confidence_quantization}
\end{equation}
The quantization intervals are $[0,0.2)$, $[0.2,0.4)$,
$[0.4,0.6)$, $[0.6,0.8)$, and $[0.8,1]$, corresponding
to levels 1 through 5, respectively.

Channel quality is obtained by first averaging the actual per-RX SINRs in the linear domain and subsequently converting the resulting mean to decibels. Specifically, the channel-quality statistic is defined as

\begin{equation}
\bar{\gamma}_{t,\mathrm{dB}}
=
10\log_{10}
\left[
\max
\left(
\frac{1}{N_{\mathrm{RX}}}
\sum_{r=1}^{N_{\mathrm{RX}}}
\gamma_{r,t}^{\mathrm{act}},
10^{-12}
\right)
\right],
\label{eq:mean_sinr_db}
\end{equation}

where $\gamma_{r,t}^{\mathrm{act}}$ is the actual linear SINR of RX $r$
at iteration $t$. Thus, $\bar{\gamma}_{t,\mathrm{dB}}$ is the dB value
of the mean linear SINR and not the arithmetic mean of the individual
SINRs expressed in dB. This quantity is represented using three
discrete channel-quality states:

\begin{equation}
q_t^{\mathrm{ch}}
=
\begin{cases}
1, & \bar{\gamma}_{t,\mathrm{dB}} < 0~\mathrm{dB},\\
2, & 0 \leq \bar{\gamma}_{t,\mathrm{dB}} < 10~\mathrm{dB},\\
3, & \bar{\gamma}_{t,\mathrm{dB}} \geq 10~\mathrm{dB}.
\end{cases}
\label{eq:channel_quality_state}
\end{equation}

The normalized separation between the two transmit-beam indices is

\begin{equation}
d_t^{\theta}
=
\frac{
\left|
\theta_{1,t}^{\mathrm{idx}}
-
\theta_{2,t}^{\mathrm{idx}}
\right|
}{
N_{\theta}-1
},
\label{eq:beam_separation}
\end{equation}

Adjacent codebook beams are separated by $5^{\circ}$.
Thus, the angular separation between the two selected TX beams is $5^{\circ}|\theta_{1,t}^{\mathrm{idx}}
-\theta_{2,t}^{\mathrm{idx}}|$.
The normalized beam-index separation is quantized into three levels represented by $q_t^{\Delta\theta}$.

Since each transmitter has four available power levels, the previous
power-pair component $q_t^{P}$ has $4^2=16$ possible values. The total
number of discrete RL states is therefore

\begin{equation}
|\mathcal{S}|
=
\binom{N_{\mathrm{RX}}}{N_{\mathrm{Att}}}
N_{\mathrm{conf}}
N_{\mathrm{ch}}
N_P^2
N_{\Delta\theta},
\label{eq:state_space_size}
\end{equation}

where

\begin{equation}
N_{\mathrm{conf}}=5,
\qquad
N_{\mathrm{ch}}=3,
\qquad
N_P=4,
\qquad
N_{\Delta\theta}=3.
\label{eq:state_dimensions}
\end{equation}

The RL action selected at iteration $t$ jointly specifies the transmit
beam and transmit power of the two transmitters:

\begin{equation}
a_t
=
\left(
\theta_{1,t},
P_{1,t},
\theta_{2,t},
P_{2,t}
\right).
\label{eq:joint_action}
\end{equation}

For each transmitter $i\in\{1,2\}$, the transmit-beam index is selected
from the 25-direction angular codebook,

\begin{equation}
\theta_{i,t}\in\Theta,
\qquad
|\Theta|=25,
\label{eq:tx_beam_codebook}
\end{equation}

while the transmit-power level satisfies

\begin{equation}
P_{i,t}
\in
\{15,20,25,30\}~\mathrm{dBm}.
\label{eq:tx_power_levels}
\end{equation}

Each receiver also selects its receive-beam direction from the same
25-direction angular codebook:

\begin{equation}
\theta_{r,t}^{\mathrm{RX}}\in\Theta,
\qquad
|\Theta|=25,
\qquad
r\in\{1,\ldots,N_{\mathrm{RX}}\}.
\label{eq:rx_beam_codebook}
\end{equation}

For BGQL and SARSA, each RX beam index is initially sampled
independently and uniformly from the 25-beam codebook.
During exploration, the intended RX beam indices may be
resampled independently and uniformly from the same codebook;
otherwise, their previous values are retained.
The RX beam indices are not independent components of the RL
action in Eq.~\eqref{eq:joint_action}. Consequently, the size of the RL action space is determined only by the joint TX beam-power decisions. Each transmitter has $25\times4=100$ beam-power combinations, and the
complete joint RL action space therefore contains

\begin{equation}
|\mathcal{A}|
=
100^2
=
10\,000.
\label{eq:action_space_size}
\end{equation}
\subsection{Multi-Objective Reward}
\label{subsec:reward}

The proposed reward combines an aggregate full-band link-rate score, a radiated transmit-power penalty, and a belief-concentration surrogate. To normalize the link-rate contribution across the considered scenarios,
a scenario-specific reference link rate $C_{\mathrm{ref}}$ is determined before RL training.

For each scenario, $M_{\mathrm{cal}}=300$ random calibration
configurations are generated. In each calibration configuration, the
two transmit-beam indices and all intended receive-beam indices are
drawn independently and uniformly from the 25-element beam codebook.
The transmit-power level of each transmitter is drawn independently and
uniformly from $\{15,20,25,30\}$~dBm. The
$N_{\mathrm{Att}}$ attacker indices are sampled uniformly without
replacement from the $N_{\mathrm{RX}}$ receiver indices. Conditional
on its intended beam index, each selected attacker is assigned an
adversarial receive-beam index drawn uniformly from the remaining 24
beam indices. The calibration link rate
$C_{\mathrm{cal}}^{(m)}$ is then evaluated using the attacked network
model, including the spatial jamming contribution.

Let $\mathcal{C}_{\mathrm{valid}}$ denote the set of positive and finite calibration link rates:

\begin{equation}
\begin{split}
\mathcal{C}_{\mathrm{valid}}=\bigl\{C_{\mathrm{cal}}^{(m)} :\;& C_{\mathrm{cal}}^{(m)}>0,\\
& C_{\mathrm{cal}}^{(m)}\ \text{is finite},\\
& m=1,\ldots,300\bigr\}.
\end{split}
\label{eq:valid_calibration_capacities}
\end{equation}

The reference link rate is defined as

\begin{equation}
C_{\mathrm{ref}}
=
\max
\left\{
Q_{0.90}\!\left(\mathcal{C}_{\mathrm{valid}}\right),
C_{\mathrm{debug}},
1~\mathrm{bit/s}
\right\},
\label{eq:capacity_reference}
\end{equation}

where $Q_{0.90}$ denotes the empirical 90th percentile and
$C_{\mathrm{debug}}$ is the aggregate link rate obtained by assigning
the central beam-codebook entry to both transmitters and all receivers,
setting both transmitters to the maximum available transmit-power level
of $30$~dBm, and evaluating the attack-free network. If
$\mathcal{C}_{\mathrm{valid}}$ is empty, the fallback rule

\begin{equation}
C_{\mathrm{ref}}
=
\max
\left\{
C_{\mathrm{debug}},
1~\mathrm{bit/s}
\right\}
\label{eq:capacity_reference_fallback}
\end{equation}

is used.

Accordingly, the normalized link-rate contribution is expressed as
\begin{equation}
R_{\mathrm{rate}}
=
\min
\left(
\frac{
\log\left(1+C_{\mathrm{tot}}^{\mathrm{act}}/C_0\right)
}{
\log\left(1+C_{\mathrm{ref}}/C_0\right)
},
1.5
\right),
\qquad
C_0=1~\mathrm{bit/s}.
\label{eq:rate_reward}
\end{equation}

The logarithmic normalization limits the influence of exceptionally large link-rate values, while the upper bound prevents the link-rate contribution
from dominating the remaining reward components. To discourage unnecessarily
high transmission power, the aggregate radiated power is normalized over its
feasible range as
\begin{equation}
R_{\mathrm{power}}
=
\left[
\frac{
P_{\mathrm{TX,tot}}-P_{\min}
}{
P_{\max}-P_{\min}+\varepsilon
}
\right]_{0}^{1},
\label{eq:power_reward}
\end{equation}
where the minimum and maximum aggregate transmit-power levels are,
respectively,
\begin{equation}
\mathcal{P}_{\mathrm{W}}
=
\left\{
10^{(p-30)/10}~\mathrm{W}:p\in\mathcal{P}
\right\},
\end{equation}
\begin{equation}
P_{\min}=2\min(\mathcal{P}_{\mathrm{W}}),
\qquad
P_{\max}=2\max(\mathcal{P}_{\mathrm{W}}).
\label{eq:minimum_and_maximum_power}
\end{equation}

Here, $\mathcal{P}_{\mathrm{W}}$ denotes the admissible transmit-power levels
after conversion from dBm to watts. This normalization assigns zero cost to
the minimum feasible aggregate transmit power and unit cost to the maximum.

The security contribution is constructed exclusively from the estimated
attacker-belief vector and does not require knowledge of the active attacker
identities. Using the top-set belief mass $B_t^{\mathrm{top}}$ defined in
Eq.~\eqref{eq:belief_confidence}, the uniform-belief reference is
\begin{equation}
B_{\mathrm{unif}}
=
\frac{N_{\mathrm{Att}}}{N_{\mathrm{RX}}}.
\label{eq:uniform_belief_mass}
\end{equation}

The belief-confidence contribution is defined as
\begin{equation}
R_{\mathrm{belief}}(t)
=
\left[
\frac{
B_t^{\mathrm{top}}-B_{\mathrm{unif}}
}{
1-B_{\mathrm{unif}}+\varepsilon
}
\right]_{0}^{1}.
\label{eq:belief_reward}
\end{equation}

A uniform belief distribution yields $R_{\mathrm{belief}}(t)=0$, whereas a more concentrated belief distribution produces a larger positive value. This contribution is calculated entirely from the anomaly-derived belief vector and does not use the identities of the active attackers. This belief-concentration surrogate does not measure whether the selected attacker identities are correct: a concentrated belief can assign high confidence to an incorrect candidate set.

The overall reward at iteration $t$ combines the link-rate, transmit-power, and attacker-localization objectives according to
\begin{equation}
R_t
=
w_R R_{\mathrm{rate}}
-
w_P R_{\mathrm{power}}
+
w_B R_{\mathrm{belief}}(t),
\label{eq:total_reward}
\end{equation}
where the weighting coefficients are selected as
$w_R=0.30$, $w_P=0.10$, and $w_B=0.60$. These values assign the largest
contribution to the estimated belief confidence while retaining explicit
incentives for link-rate enhancement and transmit-power reduction. None of
the reward components depends on the active attacker identities.

As an additional numerical safeguard, the resulting reward is bounded as
\begin{equation}
R_t
\leftarrow
\max\left(-2,\min(2,R_t)\right).
\label{eq:reward_clipping}
\end{equation}

The complete reward is calculated from the observed communication
performance, selected transmit powers, and anomaly-derived attacker belief.
The active attacker indices are not used during RL learning or action
selection.

The proposed RL framework does not employ supervised learning or a separate
labeled pre-training stage. BGQL and SARSA learn their beam-power policies
through direct interaction with the simulated network environment over the
specified learning horizon. The anomaly detector constructs the attacker
belief from the observed deviations between the DT reference and the actual
network response, without receiving labeled attacker identities. The active
attacker set is used only by the simulator to generate the adversarial
conditions and, after learning, to calculate the reported detection-accuracy
metric.

\subsection{Proposed Belief-Guided Q-Learning}
\label{subsec:q_learning}

The proposed method extends conventional tabular Q-learning by
incorporating the estimated attacker belief into its exploration
mechanism. The resulting method is referred to as belief-guided
Q-learning (BGQL). In addition to learning long-term beam-power
values, BGQL uses the accumulated attacker belief to guide exploratory
beam selection. Its off-policy update is

\begin{equation}
\begin{split}
Q(s_t,a_t)
&\leftarrow Q(s_t,a_t)
+
\alpha
\Big[
R_t
\\
&{}+\gamma
\max_{a'}
Q(s_{t+1},a')
-
Q(s_t,a_t)
\Big],
\end{split}
\label{eq:q_learning_update}
\end{equation}

where the learning rate and discount factor are

\begin{equation}
\alpha=0.3,
\qquad
\gamma=0.9.
\label{eq:rl_parameters}
\end{equation}

Exploration follows a decaying $\epsilon$ schedule, beginning at

\begin{equation}
\epsilon_0=0.30,
\label{eq:initial_epsilon}
\end{equation}

and decreasing to

\begin{equation}
\epsilon_{\min}=0.05.
\label{eq:minimum_epsilon}
\end{equation}

For a training horizon of $T=20\,000$ iterations, the exploration
probability at iteration $t$ is

\begin{equation}
\epsilon_t
=
\max
\left(
\epsilon_{\min},
\epsilon_0\delta_{\epsilon}^{\,t}
\right),
\label{eq:epsilon_schedule}
\end{equation}

where the decay factor is

\begin{equation}
\delta_{\epsilon}
=
\left(
\frac{\epsilon_{\min}}{\epsilon_0}
\right)^{1/T}.
\label{eq:epsilon_decay}
\end{equation}

During exploration, BGQL uses the current attacker belief to guide the selection of the transmit-beam pair. Each receiver is initially assigned
the weight

\begin{equation}
w_r=1-b_r,
\label{eq:belief_weight}
\end{equation}

where $b_r$ denotes the current attacker belief associated with receiver
$r$. If $b_r$ exceeds the threshold $0.20$, the corresponding weight is
replaced by

\begin{equation}
w_r=-5b_r.
\label{eq:suspicious_weight}
\end{equation}
Negative weights penalize link rate allocated to highly suspicious receivers, guiding exploratory beam selection away from serving potential attackers.

Consequently, receivers with low attacker beliefs contribute positively
to the exploratory beam-selection utility, whereas receivers considered
highly suspicious receive negative weights. For a candidate
transmit-beam pair $(\theta_1,\theta_2)$, the corresponding utility is

\begin{equation}
U(\theta_1,\theta_2)
=
\sum_{r=1}^{N_{\mathrm{RX}}}
w_r
C_r(\theta_1,\theta_2).
\label{eq:belief_guided_utility}
\end{equation}

The transmit-beam pair maximizing
Eq.~\eqref{eq:belief_guided_utility} is selected during the
belief-guided exploration step, whereas the transmit-power levels are
sampled randomly from the available power set. During exploitation,
BGQL selects an action with the largest Q-value, with random tie-breaking
when multiple actions share the same maximum value. In this manner,
accumulated security evidence influences exploratory beam selection
without using the active attacker identities in the decision policy.

Belief-guided exploration is a component of the proposed BGQL method
rather than a generic property of conventional Q-learning.
Consequently, the comparison with SARSA evaluates the complete proposed
BGQL framework against a conventional on-policy RL baseline and should
not be interpreted as an isolated comparison between the Q-learning and
SARSA update equations.

\subsection{Benchmark Schemes}
\label{subsec:baselines}

Three comparison methods are considered: SARSA, Greedy, and Random.

SARSA uses the same state representation and joint beam-power action space as Q-learning but follows the on-policy update

\begin{equation}
\begin{split}
Q(s_t,a_t)
&\leftarrow Q(s_t,a_t)
+
\alpha
\Big[
R_t
\\
&{}+\gamma Q(s_{t+1},a_{t+1})
-
Q(s_t,a_t)
\Big].
\end{split}
\label{eq:sarsa_update}
\end{equation}

During SARSA exploration, the beam and transmit-power indices are randomly sampled. During exploitation, an action is selected among those with the largest Q-value, using random tie-breaking when necessary.

The Greedy benchmark uses a non-learning local search to obtain a nominal beam-power configuration before the attacked operation. Its immediate utility is

\begin{equation}
U_{\mathrm{Greedy}}
=
w_RR_{\mathrm{rate}}
-
w_PR_{\mathrm{power}}.
\label{eq:greedy_utility}
\end{equation}

No future value, detection reward, or attacker belief is included in this optimization. Once selected, the Greedy beam-power configuration remains fixed during the attacked run.

Attacker detection for Greedy is based only on instantaneous SINR degradation,

\begin{equation}
S_r^{\mathrm{Greedy}}
=
\max
\left(
\frac{
\gamma_r^{\mathrm{exp}}-\gamma_r^{\mathrm{act}}
}{
|\gamma_r^{\mathrm{exp}}|+\varepsilon
},
0
\right).
\label{eq:greedy_detection}
\end{equation}

The $N_{\mathrm{Att}}$ receivers with the largest instantaneous values of $S_r^{\mathrm{Greedy}}$ are selected. Unlike the multi-feature detector, Greedy does not use spatial attacker-footprint matching or temporal evidence aggregation.

The Random benchmark performs no learning or optimization. At every iteration, Random independently samples the two TX beam
indices and transmit-power levels from the same joint action space used
by BGQL and SARSA. Attacker estimates are also selected randomly, providing an uninformed reference for both communication and detection performance.

Each scenario is evaluated over $30$ independent Monte Carlo
realizations with $20\,000$ iterations per realization. The attacker
set changes across realizations but is kept identical for BGQL, SARSA,
Greedy, and Random within the same realization.

After training, the final greedy BGQL and SARSA policies are
additionally evaluated over $M_{\mathrm{eval}}=100$ independently
generated attacker receive-beam realizations. Let
$b_{a}^{\mathrm{int}}$ denote the intended beam index of attacker $a$.
For evaluation realization $m$, its adversarial receive-beam index is
sampled according to

\begin{equation}
\widetilde{b}_{a}^{(m)}
\sim
\operatorname{Unif}
\left(
\{1,\ldots,25\}
\setminus
\{b_{a}^{\mathrm{int}}\}
\right),
\qquad
m=1,\ldots,100.
\label{eq:attacker_beam_sampling}
\end{equation}

Thus, conditional on the intended receive beam, each of the other 24
beam indices is selected with probability $1/24$. Because the beam
codebook is $\Theta=\{-60^\circ,-55^\circ,\ldots,60^\circ\}$, this is
equivalent to uniformly selecting one codebook direction other than the intended direction. The draws are repeated independently across evaluation realizations and attacker indices, conditional on the intended receive beams. The reported final-policy evaluation value is the arithmetic mean over these 100 attacked-policy evaluations. These post-training averages are used for the reported best-observed BGQL and SARSA configurations. In contrast, the main rate, power, and rate-to-device-power comparisons and their statistical tests use per-run averages over the last 4,000 iterations of each 20,000-iteration run,
aggregated across 30 Monte Carlo runs.

\subsection{State-Action and Exploration Complexity}
\label{subsec:complexity}

From Eqs.~\eqref{eq:state_space_size} and
\eqref{eq:action_space_size}, the number of tabular state-action entries is

\begin{equation}
|\mathcal{S}||\mathcal{A}|
=
7.2\times10^{6}
\binom{N_{\mathrm{RX}}}{N_{\mathrm{Att}}},
\label{eq:state_action_complexity}
\end{equation}

which equals $2.88\times10^{7}$, $4.32\times10^{7}$,
$1.08\times10^{8}$, and $2.016\times10^{8}$ entries in the four
scenarios, respectively. A dense table containing 64-bit values would
require approximately $0.230$, $0.346$, $0.864$, and $1.613$~GB,
respectively. In addition to the computational costs common to both RL
schemes, each BGQL exploratory step evaluates
$N_{\theta}^{2}=625$ candidate beam pairs. Its score-evaluation complexity
is therefore $\mathcal{O}(N_{\mathrm{RX}}N_{\theta}^{2})$, whereas SARSA
samples only one exploratory action.

The standard tabular Q-learning convergence theorem requires a Markov state
representation and repeated sampling of all relevant state-action pairs.
These conditions are not established in the present framework because the
state is an aggregated representation and belief-guided exploration does
not guarantee exhaustive state-action visitation. Therefore, no claim of
convergence to an optimal policy or achievement of a global optimum is made.

\section{Implementation and Simulation Setup}
\label{sec:implementation}

This section describes the implementation and simulation setup used to evaluate the proposed framework. Since the network model, propagation model, and decision-making methods have been introduced in the preceding sections, the focus here is on the simulation environment, antenna patterns, network configurations, and geographical locations of the network nodes. The main system and simulation parameters are summarized in Table~\ref{tab:system_parameters}.
\begin{figure*}[!t]
    \centering
    \includegraphics[
        width=\textwidth,
        height=0.65\textheight,
        keepaspectratio
    ]{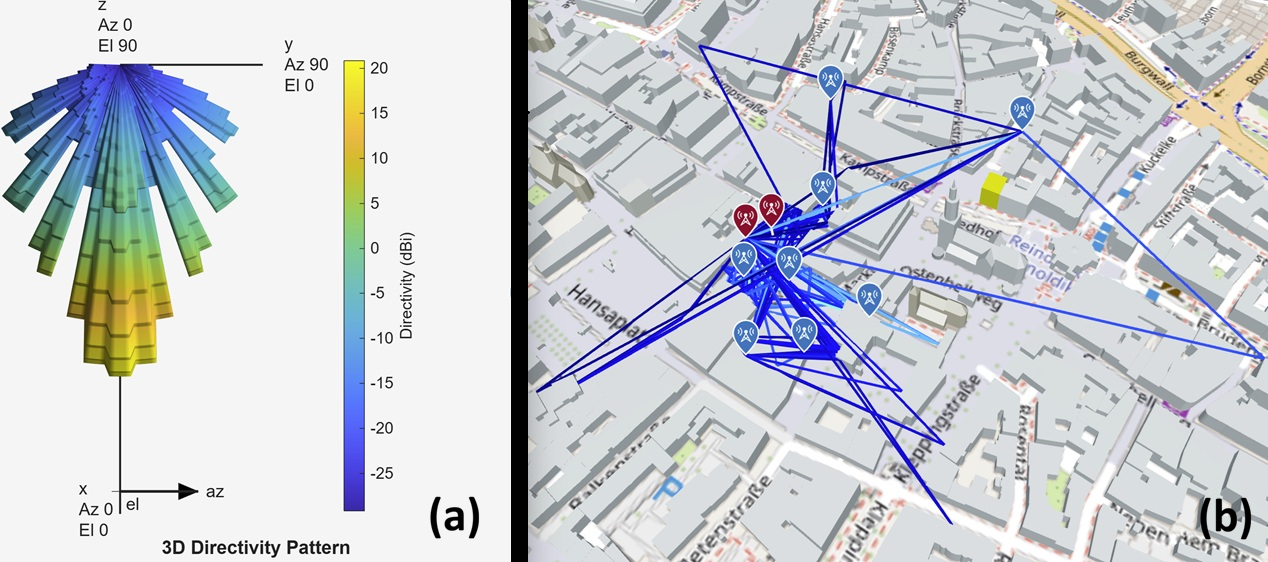}
\caption{Three-dimensional antenna directivity patterns and the site-specific ray-tracing environment used in the simulations.
Urban geometry data: \copyright~OpenStreetMap contributors
(\protect\url{https://www.openstreetmap.org/copyright}).}
    \label{fig:Fig1}
\end{figure*}

Figure~\ref{fig:Fig1} illustrates two main components of the simulation environment. Figure~\ref{fig:Fig1}(a) shows the three-dimensional antenna directivity patterns used for beamforming. The directional responses corresponding to the candidate beams are precomputed and provided to the simulation, and the appropriate pattern is selected according to the beam index during operation. Thus, the figure provides a visual representation of the directional antenna responses used in the beam-selection process rather than assuming a simplified omnidirectional antenna model.

Figure~\ref{fig:Fig1}(b) illustrates a representative network deployment in the site-specific urban environment of Dortmund used for the simulations, including the TX and RX locations and the corresponding ray-tracing propagation paths overlaid on the urban map. This representation illustrates the physical environment in which the propagation model introduced in Section~\ref{sec:system_model} is applied. The mathematical details of channel construction and propagation modeling are therefore not repeated here. The geographical locations of the network nodes are reported in Table~\ref{tab:node_locations}.

\begin{figure*}[!t]
    \centering
    \includegraphics[width=\textwidth]{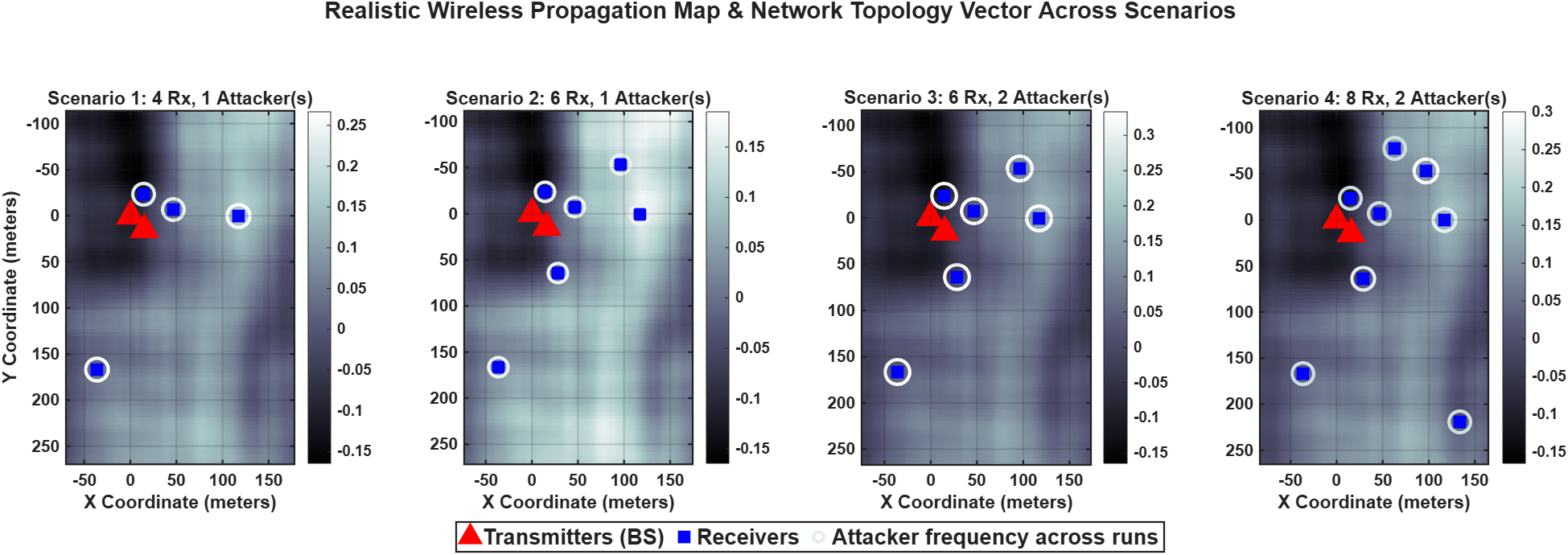}
    \caption{Spatial network configurations for the four evaluation scenarios.}
    \label{fig:Fig2}
\end{figure*}

To evaluate the proposed framework under different network conditions, four network configurations are considered, as illustrated in Figure~\ref{fig:Fig2}. The corresponding 
$(N_{\mathrm{RX}},N_{\mathrm{Att}})$ configurations are $(4,1)$, $(6,1)$, 
$(6,2)$, and $(8,2)$, respectively, with two transmitters used in all scenarios. This means that one of the four RXs is randomly selected as the attacker in the first scenario, one of the six RXs in the second scenario, two of the six RXs in the third scenario, and two of the eight RXs in the fourth scenario. These configurations enable the evaluation of the effects of increasing both the number of receivers and the number of attackers, while the fourth scenario represents the densest network configuration considered in this implementation.

Figure~\ref{fig:Fig2} shows the transmitter and receiver locations together with the spatial characteristics of the propagation environment. Since the attacker set varies across independent realizations, a particular receiver does not necessarily act as an attacker in every realization. However, 
within each realization, the same attacker set is maintained for BGQL, SARSA, Greedy, and Random, ensuring that all methods are evaluated under identical attack conditions. Consequently, the algorithms are compared using a common network topology and common threat conditions.

To characterize the network geometry and facilitate reproducibility of the simulation setup, the geographical locations of the transmitters and receivers 
are reported in Table~\ref{tab:node_locations}. For each node, the latitude, longitude, and antenna height are provided. The table illustrates one generated deployment.
For each scenario, the selected node coordinates remain
fixed across the 30 Monte Carlo runs.

\begingroup
\renewcommand{\arraystretch}{1.15}
\setlength{\tabcolsep}{6pt}

\begin{table*}[!t]
\centering\footnotesize
\caption{Geographical locations and antenna heights of the network nodes.}
\label{tab:node_locations}
\begin{tabular}{@{\extracolsep{\fill}} l l c c c @{}}
\hline
\textbf{Node} &
\textbf{Type} &
\textbf{Latitude [$^\circ$N]} &
\textbf{Longitude [$^\circ$E]} &
\textbf{Height [m]} \\
\hline

TX 1 & Transmitter & 51.513900 & 7.465300 & 15.0 \\
TX 2 & Transmitter & 51.514034 & 7.465526 & 15.0 \\
RX 1 & Receiver    & 51.513837 & 7.465973 & 1.5 \\
RX 2 & Receiver    & 51.515398 & 7.464775 & 1.5 \\
RX 3 & Receiver    & 51.513689 & 7.465509 & 1.5 \\
RX 4 & Receiver    & 51.513902 & 7.466994 & 1.5 \\
RX 5 & Receiver    & 51.514474 & 7.465712 & 1.5 \\
RX 6 & Receiver    & 51.513423 & 7.466693 & 1.5 \\
RX 7 & Receiver    & 51.513203 & 7.466202 & 1.5 \\
RX 8 & Receiver    & 51.515866 & 7.467228 & 1.5 \\

\hline
\end{tabular}
\end{table*}
\endgroup

\section{Results and Discussion} \label{sec:results}
\begin{figure*}[!t]
    \centering
    \includegraphics[width=\textwidth]{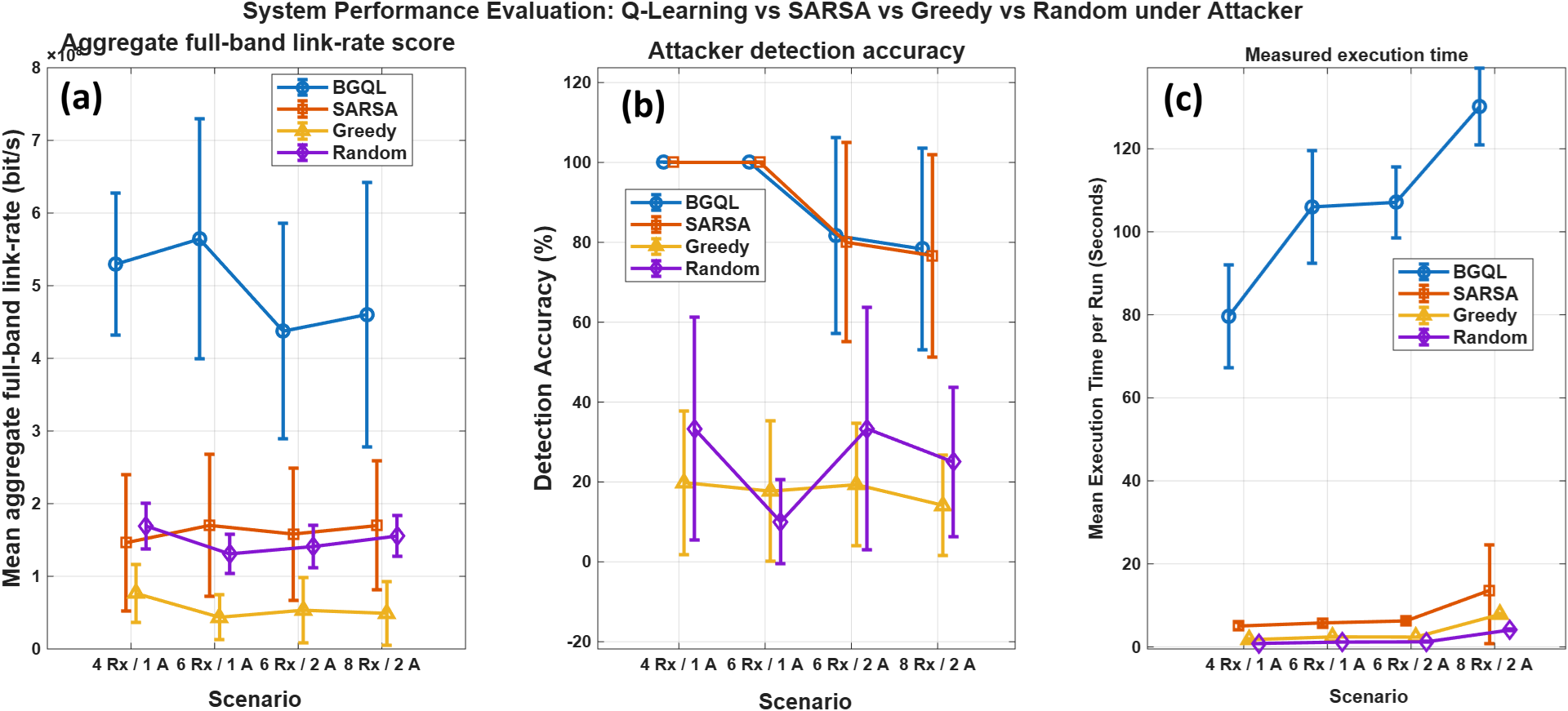}
    \caption{Performance comparison of BGQL, SARSA, Greedy, and Random across the four network scenarios: (a) aggregate full-band link-rate score; (b) attacker detection accuracy; and (c) execution time.}
    \label{fig:Fig3}
\end{figure*}

Figure~\ref{fig:Fig3} compares the performance of BGQL,  SARSA, Greedy, and Random across the four considered scenarios in terms of aggregate full-band link-rate score, attacker detection accuracy, and execution time. 
The results reveal a clear trade-off between communication, security performance and computational cost. BGQL achieves the highest aggregate full-band link-rate score across all four scenarios and, together with SARSA, provides the highest attacker detection accuracy. However, these performance gains come at the cost of considerably higher execution time.

Figure~\ref{fig:Fig3}(a) presents the average total link rate. For BGQL, the average link rates in the
$(4\,\mathrm{RX},1\,\mathrm{A})$,
$(6\,\mathrm{RX},1\,\mathrm{A})$,
$(6\,\mathrm{RX},2\,\mathrm{A})$, and
$(8\,\mathrm{RX},2\,\mathrm{A})$ scenarios are approximately
$5.3\times10^{8}$, $5.6\times10^{8}$,
$4.4\times10^{8}$, and $4.6\times10^{8}$~bps, respectively.
The highest average link rate is obtained in the second scenario, reaching approximately $560$~Mbps. A direct comparison between the second and third scenarios, both involving six receivers, shows that increasing the number of attackers from one to two reduces the BGQL link rate from approximately $5.6\times10^{8}$ to
$4.4\times10^{8}$~bps, corresponding to a reduction of approximately $21\%$. When the number of receivers is increased from six to eight while maintaining two attackers, the average link rate increases slightly to approximately $4.6\times10^{8}$~bps.

BGQL maintains a substantial link rate advantage over the other methods across all four scenarios. SARSA achieves approximately $1.45\times10^{8}$, $1.70\times10^{8}$,
$1.60\times10^{8}$, and $1.70\times10^{8}$~bps, whereas Random achieves approximately $1.70\times10^{8}$,
$1.70\times10^{8}$, $1.50\times10^{8}$, and
$1.55\times10^{8}$~bps, respectively. Greedy exhibits the lowest average link rate, with values of approximately $0.80\times10^{8}$,
$0.45\times10^{8}$, $0.55\times10^{8}$, and
$0.50\times10^{8}$~bps. These results show that the link-rate
advantage of BGQL is maintained in both the single-attacker and the
more challenging two-attacker scenarios.

Figure~\ref{fig:Fig3}(b) presents the attacker-detection performance. BGQL and SARSA use the same proposed multi-feature detector, whereas Greedy uses the instantaneous SINR-degradation
detector defined in Eq.~\eqref{eq:greedy_detection}. Random selects
the attacker estimates uniformly at random. Therefore, Greedy and
Random are evaluated as complete low-complexity benchmark schemes
rather than as detector-controlled ablations.

In the first two scenarios, each containing a single attacker, both
BGQL and SARSA achieve a detection accuracy of $100\%$. The detection
performance decreases when the number of attackers increases to two.
In the third scenario, BGQL and SARSA achieve approximately $81\%$ and
$80\%$ detection accuracy, respectively, while in the fourth scenario
the corresponding values decrease slightly to approximately $78\%$
and $77\%$. Both RL-based methods therefore maintain detection
accuracies above $75\%$ in the two-attacker scenarios, with a difference
of only approximately one percentage point between them. These results
indicate that identifying multiple attackers is more challenging;
nevertheless, both RL-based methods retain relatively high detection
accuracy.

The detection performance of Greedy and Random is considerably lower.
Greedy achieves approximately $20\%$, $18\%$, $19\%$, and $14\%$
detection accuracy across the four scenarios, respectively, whereas
the corresponding values for Random are approximately $33\%$, $10\%$,
$33\%$, and $25\%$. Increasing the number of receivers therefore does
not necessarily improve the detection performance of the baseline
schemes. For example, in the presence of two attackers, the detection
accuracy of Greedy decreases from approximately $19\%$ with six
receivers to approximately $14\%$ with eight receivers. The performance
gap is particularly pronounced in the fourth scenario, where BGQL
achieves approximately $78\%$ detection accuracy, compared with
approximately $14\%$ for Greedy, corresponding to a difference of
approximately $64$ percentage points.

The relatively large error bars observed for Greedy and Random indicate substantial variability across independent network realizations and attacker selections. The variability of BGQL and SARSA also increases in the two-attacker scenarios; however, their mean detection accuracy remains markedly higher than that of Greedy and Random. These results compare the complete detection and decision-making schemes. BGQL and
SARSA combine the proposed multi-feature detector with adaptive RL-based resource allocation, whereas Greedy and Random employ simpler
detection and decision mechanisms. The benefit of the RL-based schemes therefore extends beyond link-rate improvement to the identification of malicious receivers.

Figure~\ref{fig:Fig3}(c) illustrates the computational cost associated with these performance gains. The average execution
times of BGQL are approximately $80$, $106$, $107$, and $130$~s for the four scenarios, respectively. Thus, the execution time increases
by approximately $62.5\%$ from the first to the fourth scenario. Increasing the number of receivers from four to six in the
single-attacker case increases the execution time from approximately $80$ to $106$~s. By contrast, for the two configurations with six receivers, increasing the number of attackers from one to two results in only a minor increase in execution time, from approximately $106$
to $107$~s. The highest computational cost is observed in the densest configuration, with eight receivers and two attackers, for which the
execution time reaches approximately $130$~s.

SARSA requires considerably less computational time, with average execution times of approximately $5$, $6$, $6$, and $14$~s across the
four scenarios. Greedy requires approximately $2$, $2$, $3$, and $8$~s, respectively, while Random exhibits the lowest computational cost, with execution times of approximately $1$, $1$, $1$, and $4$~s. In the densest scenario, BGQL requires approximately $130$~s, compared with approximately $14$~s for SARSA, making SARSA
approximately nine times faster.

The higher execution time of BGQL is mainly associated with the belief-guided candidate-beam evaluation performed during exploration.
In contrast, SARSA samples a single exploratory action without performing the corresponding candidate-beam search. The execution-time
difference is therefore an expected computational cost of the proposed guided-exploration mechanism. Nevertheless, this additional computational cost is accompanied by substantially higher average
link rate and energy efficiency.

The results presented in Figure~\ref{fig:Fig3} show that BGQL achieves the highest aggregate full-band link-rate score across all considered scenarios.
BGQL and SARSA provide comparable attacker-detection performance and substantially outperform Greedy and Random in this respect. BGQL and
SARSA differ in both their temporal-difference updates and their exploration mechanisms. Therefore, the observed differences represent
the performance of the complete decision schemes rather than the isolated effect of the off-policy and on-policy update rules. In the evaluated scenarios, BGQL achieves a higher aggregate full-band link-rate score but requires longer execution time, whereas SARSA achieves similar attacker detection accuracy with shorter execution time and a lower aggregate full-band link-rate score. Greedy and Random require less execution time but achieve lower attacker detection accuracy. These results indicate a trade-off among the aggregate full-band link-rate score, attacker detection accuracy, and execution time when selecting a decision-making strategy.

\begin{figure*}[!t]
    \centering
    \includegraphics[width=\textwidth]{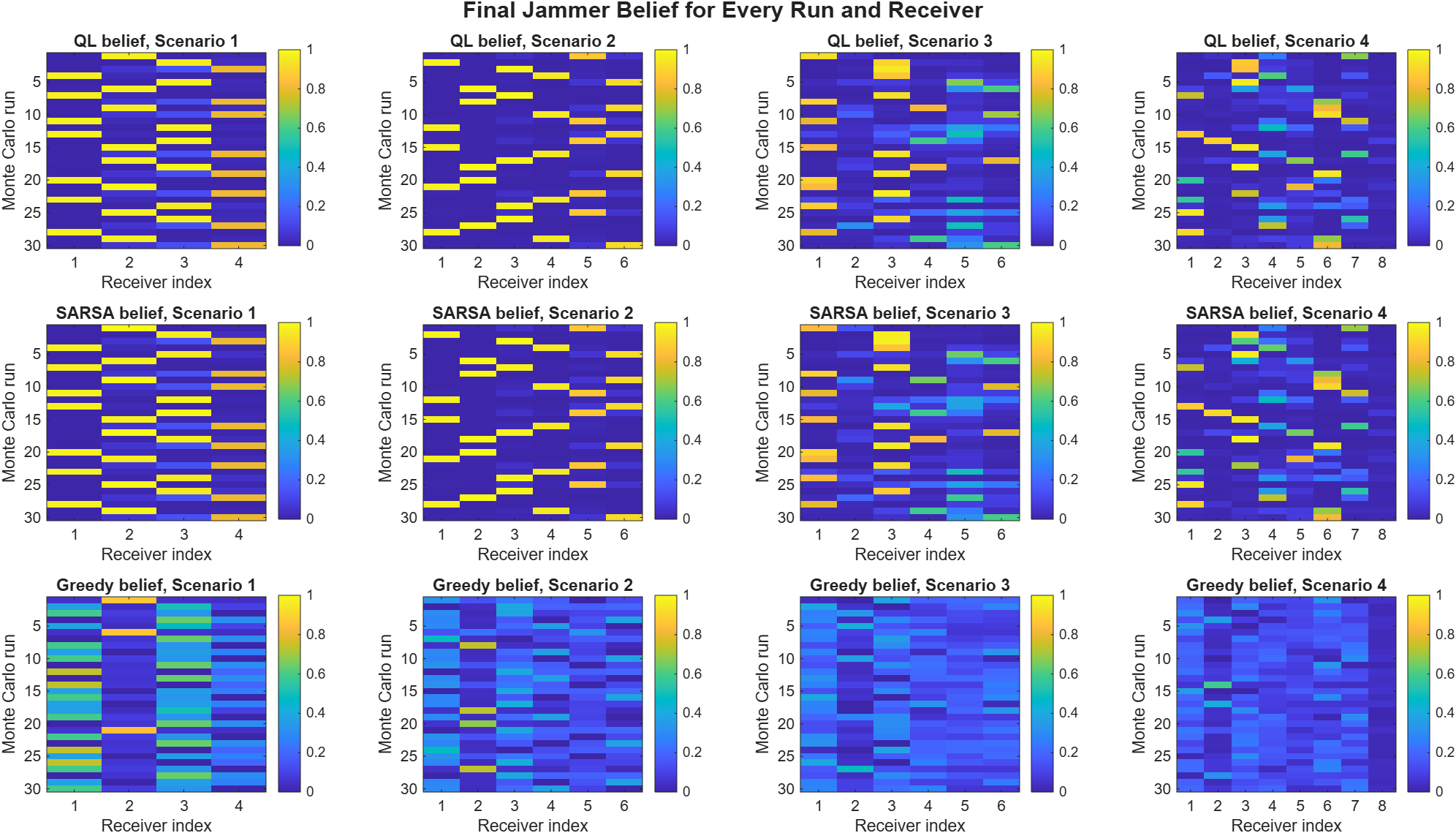}
    \caption{Receiver-level suspicion values over 30 independent Monte Carlo realizations. For BGQL and SARSA, the values represent final temporally aggregated attacker beliefs, whereas for Greedy they represent normalized instantaneous SINR-degradation scores.}
    \label{fig:Fig4}
\end{figure*}

Figure~\ref{fig:Fig4} shows the receiver-level detection scores obtained over 30 independent Monte Carlo realizations for the four network scenarios. Each column represents a receiver, while each row corresponds to an independent realization, with the color intensity indicating the score assigned to that receiver. For BGQL and SARSA, the displayed values represent the final temporally aggregated attacker beliefs, whereas for Greedy they represent the instantaneous SINR-degradation scores defined in Eq.~\eqref{eq:greedy_detection}. The figure therefore
provides a realization-level view of the separation between suspicious and normal receivers under the different detection schemes.

In the first two scenarios, where a single attacker is present, BGQL and SARSA exhibit a clear belief concentration. In most realizations, one receiver is assigned a high belief, while the beliefs of the remaining receivers stay close to zero. This behavior appears as distinct high-belief entries in the first two
scenario columns of the BGQL and SARSA rows. Since the attacker identity changes across realizations, the high-belief entries also shift among the receiver indices. This indicates that the detection mechanism is not biased toward a particular receiver but instead identifies different suspicious nodes according to the conditions of each realization. This observation is consistent with the quantitative detection results, for which both BGQL and SARSA achieve a detection accuracy of $100\%$ in the two single-attacker scenarios.

When the number of attackers increases from one to two in the third and fourth scenarios, the belief distributions become less concentrated. In contrast to the single-attacker cases, where a dominant belief close to one is frequently observed, the belief mass is distributed among multiple receivers and intermediate belief values occur more frequently. This effect is particularly evident in the fourth scenario, with eight receivers and two attackers, where the separation between suspicious and normal receivers becomes less pronounced. The same trend is reflected in the detection accuracy. BGQL and SARSA achieve approximately $81\%$ and $80\%$, respectively, in the third scenario, decreasing to approximately $78\%$ and $77\%$ in the fourth scenario. The increased dispersion of the belief values is therefore consistent with the greater difficulty of identifying multiple attackers in the larger network configurations.

The belief patterns produced by BGQL and SARSA are also notably similar. Across the four scenarios, receivers assigned high beliefs by BGQL are generally assigned similarly high beliefs by SARSA. This observation agrees with the small difference between their detection accuracies. Despite the substantial difference between BGQL and SARSA in terms of link rate and computational cost, their attacker detection performance remains comparable. The main distinction between these two RL methods is therefore observed in their communication-performance-complexity trade-off rather than in their ability to identify malicious receivers.

A different behavior is observed for Greedy. As shown in the third row
of Figure~\ref{fig:Fig4}, the instantaneous SINR-degradation scores are
more broadly distributed across the receivers, and a dominant score
associated with a particular suspicious receiver is less frequently
observed. This behavior is consistent with the considerably lower
detection accuracy of Greedy, which is approximately $20\%$, $18\%$,
$19\%$, and $14\%$ for the four scenarios, respectively. The difference
becomes particularly apparent in the fourth scenario, where the
suspicion scores are predominantly distributed over low and
intermediate levels without a clear separation between suspicious and
normal receivers. Hence, Greedy exhibits not only lower average
detection accuracy but also weaker discrimination between normal and
malicious receivers at the instantaneous-score level.

Taken together, the receiver-level detection scores in
Figure~\ref{fig:Fig4} and the aggregate detection accuracies presented
in the middle panel of Figure~\ref{fig:Fig3} provide complementary
views of the detection process. For BGQL and SARSA, the heatmaps
show the final temporally aggregated belief distributions, whereas for
Greedy they show the instantaneous SINR-degradation scores. In the
single-attacker scenarios, the concentrated belief patterns produced by
BGQL and SARSA correspond to a detection accuracy of $100\%$. In
the two-attacker scenarios, the increased dispersion of their belief
values is accompanied by a reduction in detection accuracy to
approximately $77\%$-$81\%$. In contrast, the weaker separation
between the instantaneous suspicion scores assigned to normal and
malicious receivers by Greedy is consistent with its substantially
lower detection accuracy of approximately $14\%$-$20\%$. These results
indicate that the multi-feature belief-based detector used with the RL
approaches provides stronger discrimination of malicious receivers than
the instantaneous SINR-degradation detector used with Greedy, although
identifying multiple attackers becomes more challenging as the numbers
of attackers and receivers increase.

\begin{figure*}[!t]
    \centering
    \includegraphics[width=\textwidth]{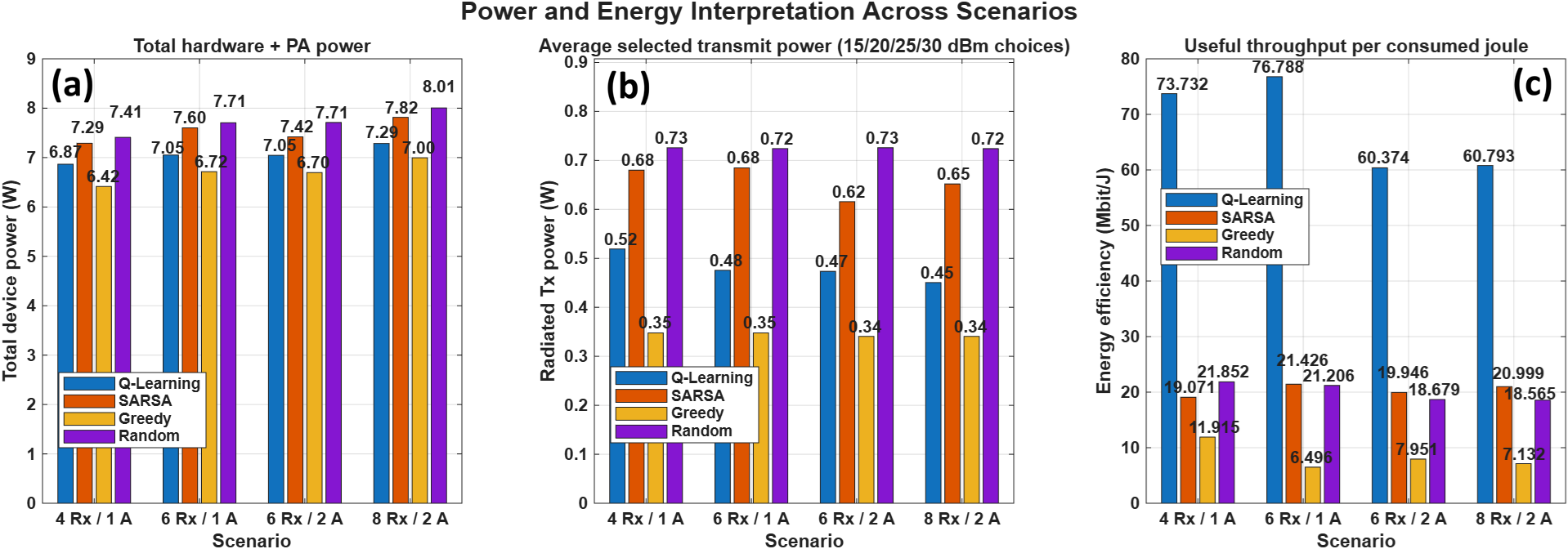}
\caption{Comparison of BGQL, SARSA, Greedy, and Random across the four evaluated fixed-layout configurations: (a) total device power consumption; (b) average aggregate selected radiated transmit power; and (c) mean rate-to-device-power score.}
    \label{fig:Fig6}
\end{figure*}

Figure~\ref{fig:Fig6} compares BGQL, SARSA, Greedy, and Random in terms of total power consumption, average selected transmit power, and energy efficiency across the four network scenarios. The results indicate that BGQL achieves a favorable balance between power consumption and communication performance, resulting in substantially higher energy efficiency than the other evaluated methods.

Figure~\ref{fig:Fig6}(a) reports the total device power consumption, including the hardware and power-amplifier components. For BGQL, the total power consumption is 6.87, 7.05, 7.05, and 7.29~W across the four scenarios, respectively. The corresponding values are 7.29, 7.60, 7.42, and 7.82~W for SARSA and 7.41, 7.71, 7.71, and 8.01~W for Random. BGQL therefore operates at a lower total power level than SARSA and Random in all four configurations. Moreover, as the network size increases from four to eight receivers, its total power consumption increases only from 6.87 to 7.29~W. Greedy exhibits the lowest absolute power consumption, with values of 6.42, 6.72, 6.70, and 7.00~W. However, this reduction in consumed power is accompanied by the substantially lower aggregate full-band link-rate score reported previously and consequently does not translate into a higher rate-to-device-power score.

Figure~\ref{fig:Fig6}(b) presents the average aggregate
selected radiated transmit power,
\begin{equation}
\overline{P}_{\mathrm{TX,tot}}
=
\frac{1}{M}
\sum_{m=1}^{M}
\left(P_{1,m}+P_{2,m}\right),
\end{equation}
where $M$ denotes the number of simulation runs. 
BGQL selects average aggregate radiated transmit powers of 0.52, 0.48,
0.47, and 0.45~W across the four configurations. By comparison, the
corresponding values for SARSA are 0.68, 0.68, 0.62, and 0.65~W, whereas
Random consistently operates at higher levels of approximately
0.72-0.73~W. Hence, the link-rate improvement achieved by BGQL is not obtained through increased transmit power. Instead, BGQL simultaneously achieves higher link rate while selecting lower transmit-power levels than SARSA and Random. The gradual reduction from 0.52~W in the first scenario to 0.45~W in the fourth further indicates that the learned policy does not rely on increasing transmit power as the network configuration becomes more demanding.

This behavior is reflected more clearly in the energy-efficiency results shown in Figure~\ref{fig:Fig6}(c). BGQL achieves 73.732, 76.788, 60.374, and 60.793~Mbit/J across the four scenarios, respectively.
These values are substantially higher than those obtained with the other methods. SARSA achieves 19.071, 21.426, 19.946, and 20.999~Mbit/J, while Random obtains 21.852, 21.206, 18.679, and 18.565~Mbit/J. Greedy yields 11.915, 6.496, 7.951, and 7.132~Mbit/J. BGQL therefore achieves the highest mean rate-to-device-power score in every considered network configuration. 
The influence of increasing attack intensity can be examined directly by comparing the second and third scenarios, for which the number of receivers remains fixed at six while the number of attackers increases from one to two. Under BGQL, the energy efficiency decreases from 76.788 to 60.374~Mbit/J, corresponding to a reduction of approximately 21.4\%.
Despite this degradation, BGQL retains a substantial energy efficiency advantage over all competing methods. When the network is further enlarged to eight receivers with two attackers, the energy efficiency remains nearly unchanged at 60.793~Mbit/J, while the average selected transmit power decreases further to 0.45~W. Within the four evaluated fixed-layout configurations, BGQL maintains a
higher mean rate-to-power score than the competing methods without relying
on a higher aggregate radiated transmit power.

These results show that the higher aggregate full-band link-rate score achieved by BGQL is not attributable to increased transmit power. Compared with SARSA and Random, BGQL achieves higher aggregate full-band link-rate and rate-to-device-power scores while operating with lower average aggregate radiated transmit power and lower total device power consumption. Although Greedy consumes less device power, its substantially lower aggregate full-band link-rate score results in a lower rate-to-device-power score. These findings support the comparative advantage of BGQL under the stated simulation model; they do not establish higher delivered network throughput or deployment-level energy efficiency.

\begin{figure*}[!t]
    \centering
    \includegraphics[width=\textwidth]{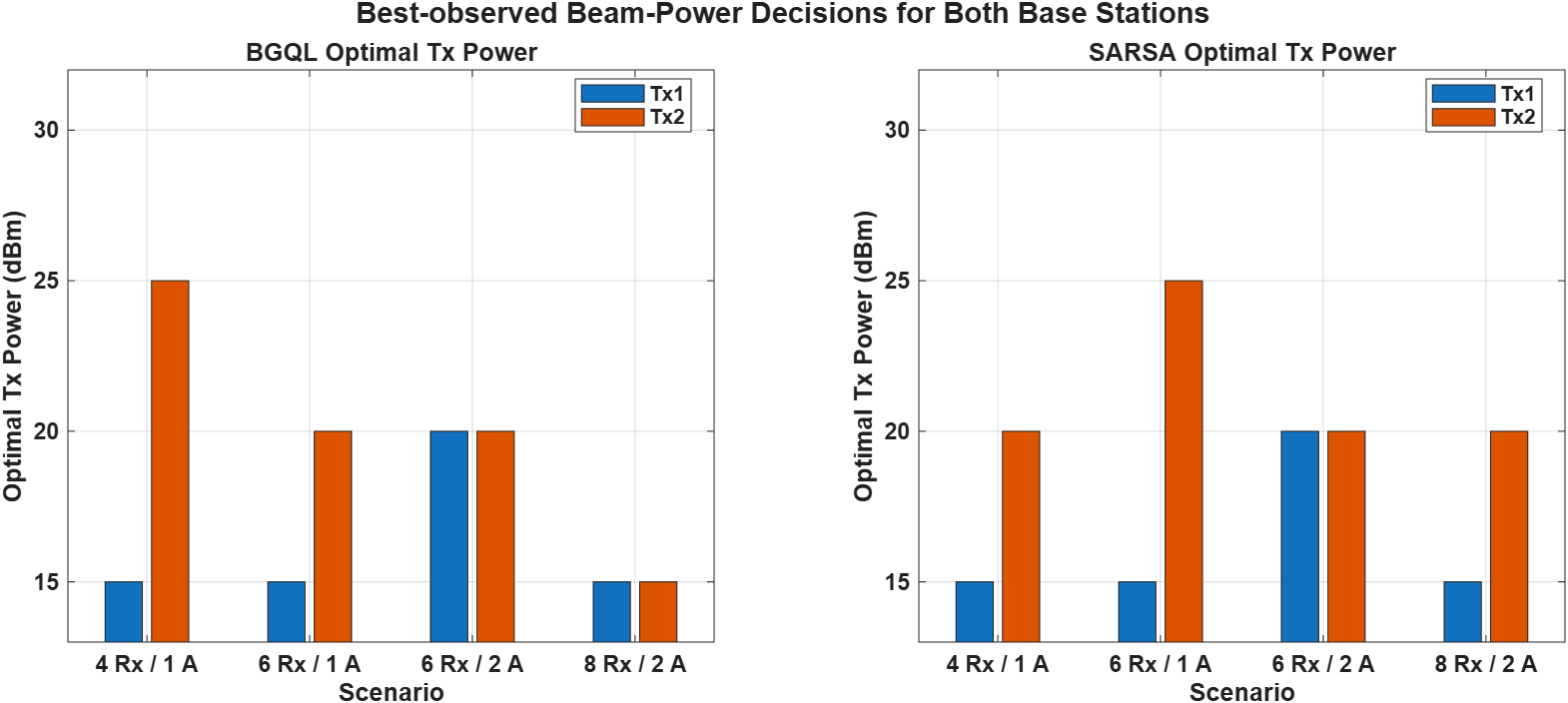}
    \caption{Best-observed TX beam-power configurations obtained by BGQL and SARSA across the four evaluated network configurations.}
    \label{fig:Fig7}
\end{figure*}

Figure~\ref{fig:Fig7} presents the best-observed beam-power decision obtained for the two base stations, TX1 and TX2, across the four network scenarios. These values correspond to the best final configuration identified by each algorithm and should be distinguished from the average transmit-power values calculated
over 30 independent Monte Carlo runs. The figure therefore illustrates the joint impact of the selected beam direction and transmit-power level on the resulting link rate.

For BGQL, the best configuration in the first scenario uses a transmit angle of $0^\circ$ for TX1 and $-60^\circ$ for TX2. The corresponding transmit powers are 15 and 25~dBm, respectively. This combination yields a link rate of $1.4022\times10^{9}$~bps, a device power consumption of 6.4154~W, and an energy efficiency of $2.1858\times10^{8}$~bit/J. In the second scenario,the best-observed transmit angles are $60^\circ$ for TX1 and $-60^\circ$ for TX2, with transmit powers of 15 and 20~dBm, respectively. The resulting link rate is $4.6093\times10^{8}$~bps. In the third scenario, BGQL selects $-30^\circ$ for TX1 and $-60^\circ$ for TX2, while both transmitters operate at 20~dBm, resulting in a link rate of $5.0090\times10^{8}$~bps.

The fourth scenario further illustrates the role of joint beam-power optimization. In this case, TX1 operates at $0^\circ$ and TX2 at $-60^\circ$, while both transmitters select the lowest available power level of 15~dBm.
Despite this low transmit-power setting, the configuration achieves a link rate of $1.9603\times10^{9}$~bps and an energy efficiency of $3.1282\times10^{8}$~bit/J. This result demonstrates that high link rate does not necessarily require higher transmit power. When the beamforming directions are properly aligned with the propagation conditions and network topology, the transmitted energy can be concentrated along more
effective propagation paths, allowing high link rate to be achieved with lower power.

SARSA also selects different beam-power combinations across the four scenarios. In the first scenario, TX1 and TX2 operate at $-25^\circ$ and $-5^\circ$, with transmit powers of 15 and 20~dBm, respectively. In the second scenario, TX1 selects $-15^\circ$ and TX2 selects $-60^\circ$, with transmit powers of 15 and 25~dBm. In the third scenario, the selected angles
are $10^\circ$ for TX1 and $45^\circ$ for TX2, while both transmitters operate at 20~dBm. In the fourth scenario, TX1 operates at $15^\circ$ with 15~dBm, whereas TX2 operates at $60^\circ$ with 20~dBm.

Although SARSA occasionally reaches very high-link-rate configurations,
its average performance over the 30 independent realizations remains
lower than that of BGQL. The average link rates of BGQL across the four
scenarios are $5.2968\times10^{8}$, $5.6424\times10^{8}$, $4.3738\times10^{8}$, and $4.6027\times10^{8}$~bps, respectively.
The corresponding SARSA values are $1.452\times10^{8}$, $1.701\times10^{8}$, $1.602\times10^{8}$, and $1.701\times10^{8}$~bps. Hence, the performance of an individual best configuration should be distinguished from the average behavior of the algorithm across independent realizations.

The results in Figure~\ref{fig:Fig7} also show that transmit-power selection cannot be considered independently of beamforming. Increasing the transmit power does not necessarily improve link rate, as the achievable link rate also depends on the selected beam directions and the resulting channel gains and interference levels. The achieved link-rate score depends on both the selected TX beams and the sampled RX beams. However, the RL action controls only the two TX beam indices and their transmit-power levels; RX beam indices
are sampled according to the procedure described in Section 3.4. The control objective is therefore to select suitable TX beam-power combinations under the resulting RX beam configurations. To complement Figure~\ref{fig:Fig7}, the receiver beam angles associated with the best-observed BGQL and SARSA configurations are summarized in Table~\ref{tab:optimal_rx_angles}.

\begin{table*}[!t]
\centering
\caption{Receiver beam angles associated with the best-observed
BGQL and SARSA configurations across the four network scenarios.}
\label{tab:optimal_rx_angles}
\resizebox{\linewidth}{!}{%
\begin{tabular}{l l c c c c c c c c}
\hline
\textbf{Scenario} & \textbf{Method} &
\textbf{RX1} & \textbf{RX2} & \textbf{RX3} & \textbf{RX4} &
\textbf{RX5} & \textbf{RX6} & \textbf{RX7} & \textbf{RX8} \\
\hline
4 RX, 1 A & BGQL
& $60^\circ$ & $0^\circ$ & $-45^\circ$ & $-60^\circ$
& -- & -- & -- & -- \\

4 RX, 1 A & SARSA
& $-15^\circ$ & $-60^\circ$ & $50^\circ$ & $-5^\circ$
& -- & -- & -- & -- \\

6 RX, 1 A & BGQL
& $25^\circ$ & $15^\circ$ & $-30^\circ$ & $-50^\circ$
& $-60^\circ$ & $-35^\circ$ & -- & -- \\

6 RX, 1 A & SARSA
& $45^\circ$ & $-20^\circ$ & $55^\circ$ & $35^\circ$
& $0^\circ$ & $30^\circ$ & -- & -- \\

6 RX, 2 A & BGQL
& $-20^\circ$ & $55^\circ$ & $40^\circ$ & $20^\circ$
& $0^\circ$ & $-10^\circ$ & -- & -- \\

6 RX, 2 A & SARSA
& $40^\circ$ & $60^\circ$ & $60^\circ$ & $-25^\circ$
& $-45^\circ$ & $-20^\circ$ & -- & -- \\

8 RX, 2 A & BGQL
& $-25^\circ$ & $35^\circ$ & $55^\circ$ & $-45^\circ$
& $-20^\circ$ & $30^\circ$ & $40^\circ$ & $-55^\circ$ \\

8 RX, 2 A & SARSA
& $-30^\circ$ & $10^\circ$ & $-25^\circ$ & $-30^\circ$
& $0^\circ$ & $-5^\circ$ & $0^\circ$ & $30^\circ$ \\
\hline
\end{tabular}
}
\end{table*}

\begin{figure*}[!t]
    \centering
    \includegraphics[width=\textwidth]{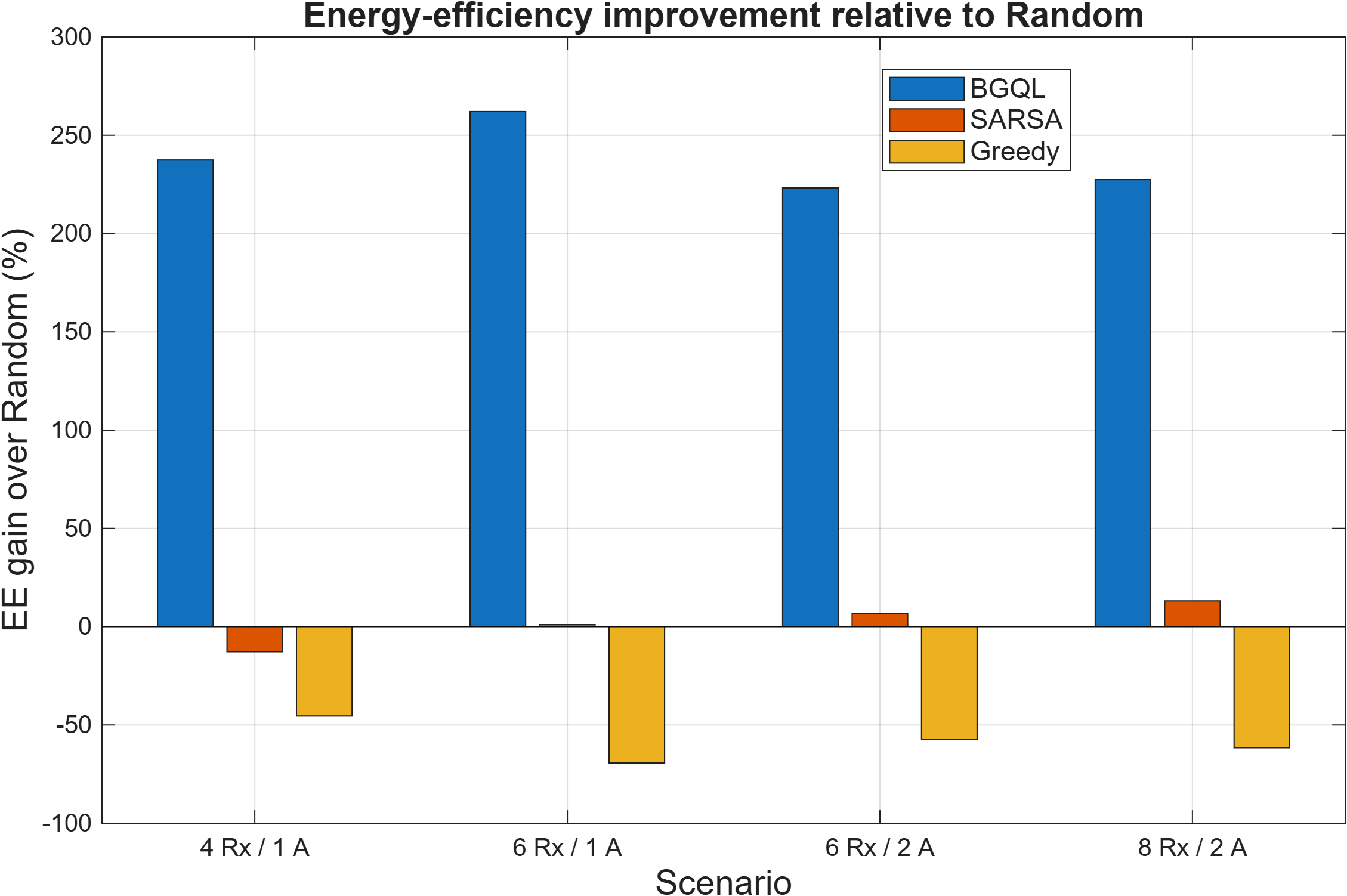}
    \caption{Energy-efficiency improvement of BGQL, SARSA, and Greedy relative to the Random baseline across the four network scenarios.}
    \label{fig:Fig8}
\end{figure*}

Figure~\ref{fig:Fig8} reports the percentage change in energy efficiency relative to the Random baseline. BGQL achieves improvements of approximately $237.4\%$, $262.1\%$, $223.2\%$, and $227.5\%$ across the four scenarios, respectively. The energy-efficiency advantage therefore remains consistently above $220\%$, including in the two-attacker scenarios.

SARSA shows a much smaller variation relative to Random. In the first scenario, its energy efficiency is approximately $12.7\%$ lower than Random. In the second scenario, it improves by about $1.0\%$, while in the third and fourth scenarios the corresponding gains are approximately $6.8\%$ and $13.1\%$. Greedy remains below the Random baseline in all four cases, with energy-efficiency reductions of approximately $45.5\%$, $69.4\%$, $57.4\%$, and $61.6\%$, respectively.

The results in Figure~\ref{fig:Fig8} are consistent with the beam-power decisions shown in Figure~\ref{fig:Fig7}. In the fourth scenario, BGQL achieves $1.9603\times10^{9}$~bps in its best configuration while operating both transmitters at only 15~dBm, with transmit angles of $0^\circ$ and $-60^\circ$. At the level of the averaged results, BGQL achieves an energy efficiency of 60.793~Mbit/J with an average transmit power of 0.451~W, whereas Random reaches only 18.565~Mbit/J despite using a higher average transmit power of 0.724~W. These results demonstrate that link rate and energy efficiency are determined by the joint selection of beam directions and transmit-power levels.

\begin{figure*}[!t]
    \centering
    \includegraphics[width=\textwidth]{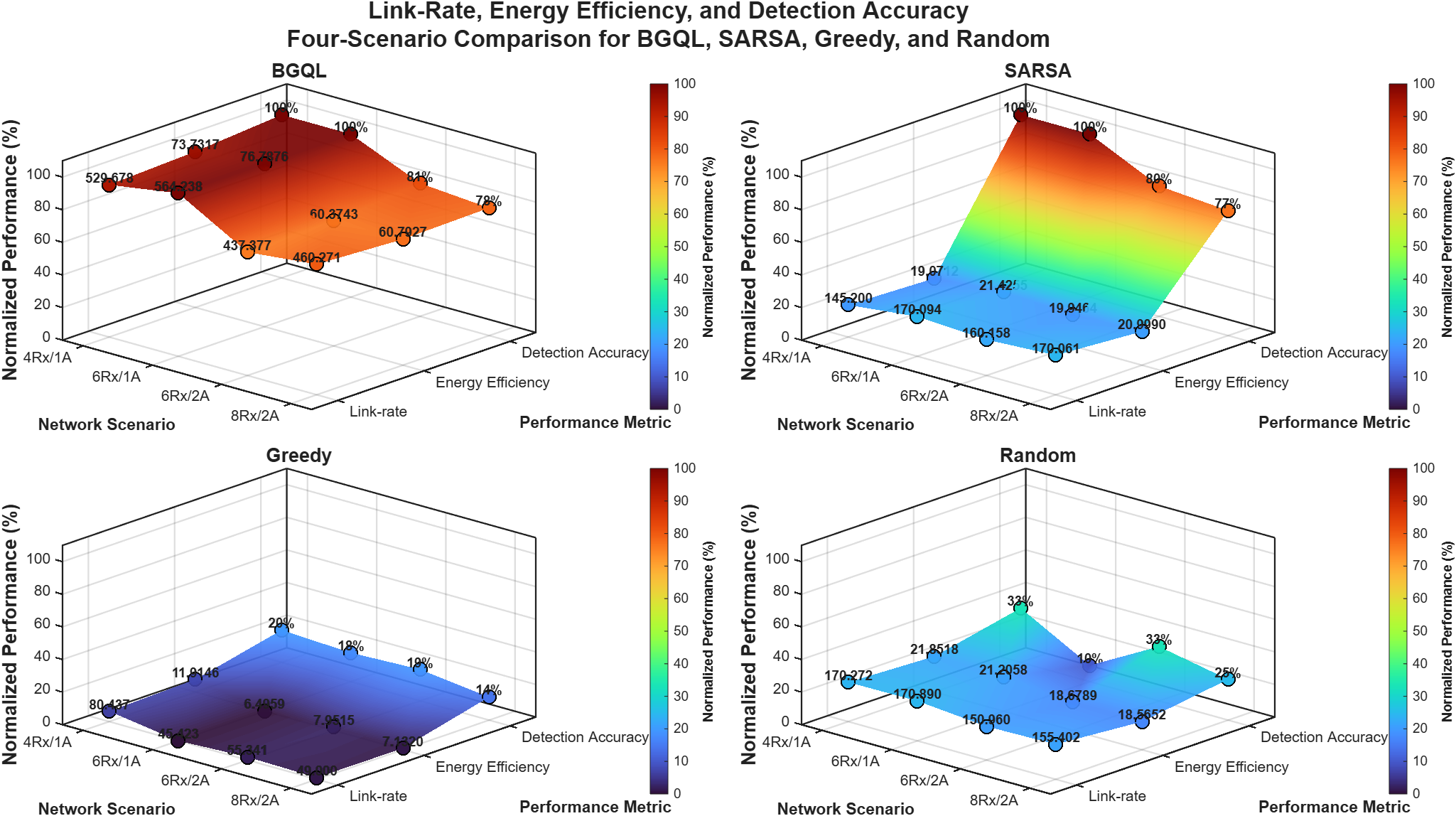}
    \caption{Joint comparison of link rate, energy efficiency, and attacker detection accuracy for BGQL, SARSA, Greedy, and Random across the four network scenarios.}
    \label{fig:Fig9}
\end{figure*}

Figure~\ref{fig:Fig9} provides an integrated comparison of BGQL, SARSA, Greedy, and Random across four network configurations in terms of link rate, energy efficiency, and attacker detection accuracy. The results demonstrate a clear overall advantage of BGQL when these three performance dimensions are considered jointly. Although SARSA achieves attacker detection accuracy comparable to that of BGQL, the substantial differences in link rate and energy efficiency indicate that BGQL maintains a high detection capability while utilizing communication and energy resources more effectively.

The average link rates achieved by BGQL in the $4\mathrm{RX}/1\mathrm{A}$, $6\mathrm{RX}/1\mathrm{A}$, $6\mathrm{RX}/2\mathrm{A}$, and $8\mathrm{RX}/2\mathrm{A}$ scenarios are 529.7, 564.2, 437.4, and 460.3~Mbps, respectively. The corresponding values for SARSA are 145.2, 170.1, 160.2, and 170.1~Mbit/s. BGQL therefore maintains a substantial link-rate advantage across all considered configurations. A direct comparison between the $6\mathrm{RX}/1\mathrm{A}$ and $6\mathrm{RX}/2\mathrm{A}$ scenarios shows that introducing a second attacker while keeping the number of receivers fixed reduces the BGQL link rate from 564.2 to 437.4~Mbps. Despite this reduction, the resulting link rate remains considerably higher than those achieved by SARSA, Greedy, and Random. Furthermore, increasing the number of receivers from six to eight in the presence of two attackers increases the link rate to 460.3~Mbps, indicating that the learned policy remains effective in utilizing the available network resources as the network becomes denser.

An even more pronounced performance gap is observed in terms of energy efficiency. BGQL achieves 73.73, 76.79, 60.37, and 60.79~Mbit/J across the four scenarios, respectively, whereas SARSA operates within a range of 19.07-21.43~Mbit/J and Random within 18.57-21.85~Mbit/J. Greedy achieves considerably lower values, ranging from 6.50 to 11.92~Mbit/J. Hence, the link-rate advantage of BGQL is not obtained at the expense of disproportionately higher energy consumption. Instead, the substantially higher ratio of useful transmitted data to consumed energy indicates more effective utilization of the available power resources. This observation is consistent with the joint beamforming and transmit-power optimization results, where coordinated selection of the beam directions and transmit-power levels enables higher link rate without relying solely on increased transmit power.

Regarding attacker detection, BGQL and SARSA both achieve $100\%$ accuracy in the two single-attacker scenarios. When the number of attackers increases to two, the detection accuracy of BGQL decreases to approximately $81\%$ and $78\%$ in the $6\mathrm{Rx}/2\mathrm{A}$ and $8\mathrm{Rx}/2\mathrm{A}$ scenarios, respectively, while SARSA achieves approximately $80\%$ and $77\%$ under the same conditions. The difference in detection performance between the two RL methods is therefore relatively small, whereas their differences in link rate and energy efficiency remain substantial. This indicates that the advantage of BGQL arises primarily from its ability to preserve a high level of attacker detection while simultaneously improving communication and energy performance, rather than from optimizing a single performance metric.

In contrast, Greedy and Random do not achieve a comparable balance across the three performance metrics. Greedy attains detection accuracies of approximately $20\%$, $18\%$, $19\%$, and $14\%$ across the four scenarios, whereas Random achieves approximately $33\%$, $10\%$, $33\%$, and $25\%$, respectively. Their lower detection accuracy is accompanied by substantially lower link rate and energy efficiency, further highlighting the performance gap between the RL-based policies and the baseline methods, particularly under more challenging network conditions.

Increasing the number of attackers from one to two reduces thelink rate, energy efficiency, and detection accuracy of BGQL; nevertheless, its relative advantage is preserved under the more challenging configurations. In the $8\mathrm{RX}/2\mathrm{A}$ scenario, BGQL achieves a link rate of 460.3~Mbps, an energy efficiency of 60.79~Mbit/J, and a detection accuracy of approximately $78\%$. Under the same configuration, SARSA achieves a comparable detection accuracy of approximately $77\%$, but provides only 170.1~Mbps of link rate and 21.0~Mbit/J of energy efficiency. Thus, BGQL provides approximately $2.7\times$ the link rate and $2.9\times$ the energy efficiency of SARSA while maintaining a comparable level of attacker detection. This result demonstrates that the advantage of BGQL under more complex network conditions extends beyond any individual performance metric.

Thus, the three performance dimensions presented in Figure~\ref{fig:Fig9} shows that BGQL achieves the highest aggregate full-band link-rate and rate-to-device-power scores among the evaluated methods in all four scenarios. Its attacker identification performance is comparable to that of SARSA and higher than that of the Greedy and Random benchmarks. These results support the comparative performance of the complete BGQL scheme within the evaluated fixed-layout, matched-model simulations.

\subsection{Statistical Analysis}
\label{subsec:statistical_analysis}

The inferential analysis evaluates the aggregate full-band link-rate score, the rate-to-device-power score, and known-cardinality attacker-identification recall using 30 matched Monte Carlo runs for each evaluated fixed-layout configuration.

For each metric, configuration, and method comparison, paired run-level differences were calculated. For comparisons whose paired differences were not all zero, the two-sided Wilcoxon signed-rank test was used to assess whether the distribution of differences is symmetric about zero. This test does not require normality but relies on independent paired differences and their approximate symmetry.

For the two single-jammer BGQL--SARSA recall comparisons, all 30 paired differences are zero. These comparisons are reported as not testable by the signed-rank procedure and are excluded from the Holm correction family. Cohen's $d_z$ is undefined because both the mean and the sample standard deviation of the paired differences are zero.

Table~\ref{tab:statistical_significance} reports raw $p$-values. Holm's step-down correction was applied across the 34 testable comparisons in the table. All comparisons with reported raw $p<0.001$ remain significant after correction, while the two BGQL--SARSA two-jammer recall comparisons remain non-significant.

For each testable comparison, the $p$-value is reported together with Cohen's paired-sample effect size,
\begin{equation}
d_z = \frac{\overline{\Delta}}{s_{\Delta}},
\label{eq:cohen_dz}
\end{equation}
where $\Delta$ is the vector of paired run-level differences, $\overline{\Delta}$ is its mean, and $s_{\Delta}$ is its sample standard deviation. A positive $d_z$ indicates that the first method in the comparison has the larger run-level mean.

For the aggregate full-band link-rate score, BGQL significantly outperforms SARSA, Greedy, and Random in all four evaluated configurations. In the direct comparison with SARSA, the raw $p$-values for 4RX/1A, 6RX/1A, 6RX/2A, and 8RX/2A are $3.41\times10^{-5}$, $7.69\times10^{-6}$, $9.71\times10^{-5}$, and $1.48\times10^{-4}$, respectively. The corresponding Cohen's $d_z$ values are 1.16, 1.36, 1.00, and 0.90. Comparisons with Greedy and Random also yield raw $p<0.001$ and large positive effect sizes in every configuration. These results provide statistical support for the link-rate advantage of BGQL.

The rate-to-device-power score exhibits a similar pattern. BGQL significantly outperforms SARSA, Greedy, and Random in all four evaluated configurations. For BGQL versus SARSA, all raw $p$-values are below 0.001, while Cohen's $d_z$ ranges from 0.95 to 1.42. The effect sizes range from 1.23 to 1.69 for comparisons with Greedy and from 1.07 to 1.50 for comparisons with Random. These significant paired differences and large positive effect sizes support the rate-to-device-power advantage of BGQL.

For attacker-identification recall, BGQL and SARSA achieve identical run-level values in the two single-jammer configurations. In the two-jammer configurations, their comparisons yield reported raw $p$-values of 1.00 and Cohen's $d_z$ values of 0.10 and 0.18, respectively, with no statistically significant difference after Holm correction.

BGQL also achieves significantly higher attacker-identification recall than Greedy and Random in all four evaluated configurations. The effect sizes range from 1.65 to 2.98 for BGQL versus Greedy and from 1.20 to 2.95 for BGQL versus Random. Together, these results demonstrate the communication and power-efficiency advantages of BGQL over SARSA alongside similar observed attacker-identification recall, as well as higher recall than the Greedy and Random baselines.

Execution time is summarized descriptively as a measure of computational cost and is not included in the inferential comparison family. BGQL has longer mean execution times than the other evaluated methods in all four configurations, reflecting the additional candidate-beam evaluations used by its belief-guided exploration mechanism. This computational cost accompanies its higher mean link-rate and rate-to-device-power scores.

Table~\ref{tab:statistical_significance} summarizes the paired statistical comparisons for the three performance metrics.

\begingroup
\footnotesize
\renewcommand{\arraystretch}{1.05}
\setlength{\tabcolsep}{5pt}

\begin{table*}[!t]
\centering\footnotesize
\caption{Paired statistical comparisons of BGQL with SARSA, Greedy, and Random across four evaluated fixed-layout configurations. Entries report raw $p$-values and Cohen's $d_z$. Holm correction is applied across the 34 testable comparisons. The two comparisons with all-zero paired differences are excluded from the correction family and reported as not testable, with undefined effect sizes.}
\label{tab:statistical_significance}
\begin{tabular}{lllcc}
\hline
\textbf{Metric} & \textbf{Scenario} & \textbf{Comparison} & \(p\)-value & Cohen's \(d_z\) \\
\hline

\shortstack[l]{Aggregate full-band\\link-rate score} & 4RX/1A & BGQL vs. SARSA & $3.41\times10^{-5}$ & 1.16 \\
& 4RX/1A & BGQL vs. Greedy & $1.97\times10^{-5}$ & 1.32 \\
& 4RX/1A & BGQL vs. Random & $3.72\times10^{-5}$ & 1.20 \\
\cline{2-5}
& 6RX/1A & BGQL vs. SARSA & $7.69\times10^{-6}$ & 1.36 \\
& 6RX/1A & BGQL vs. Greedy & $6.98\times10^{-6}$ & 1.73 \\
& 6RX/1A & BGQL vs. Random & $7.69\times10^{-6}$ & 1.44 \\
\cline{2-5}
& 6RX/2A & BGQL vs. SARSA & $9.71\times10^{-5}$ & 1.00 \\
& 6RX/2A & BGQL vs. Greedy & $1.80\times10^{-5}$ & 1.40 \\
& 6RX/2A & BGQL vs. Random & $5.31\times10^{-5}$ & 1.09 \\
\cline{2-5}
& 8RX/2A & BGQL vs. SARSA & $1.48\times10^{-4}$ & 0.90 \\
& 8RX/2A & BGQL vs. Greedy & $3.41\times10^{-5}$ & 1.25 \\
& 8RX/2A & BGQL vs. Random & $4.07\times10^{-5}$ & 1.00 \\
\hline

\shortstack[l]{Rate-to-device-power\\score} & 4RX/1A & BGQL vs. SARSA & $3.41\times10^{-5}$ & 1.23 \\
& 4RX/1A & BGQL vs. Greedy & $3.11\times10^{-5}$ & 1.26 \\
& 4RX/1A & BGQL vs. Random & $3.41\times10^{-5}$ & 1.25 \\
\cline{2-5}
& 6RX/1A & BGQL vs. SARSA & $7.69\times10^{-6}$ & 1.42 \\
& 6RX/1A & BGQL vs. Greedy & $6.98\times10^{-6}$ & 1.69 \\
& 6RX/1A & BGQL vs. Random & $7.69\times10^{-6}$ & 1.50 \\
\cline{2-5}
& 6RX/2A & BGQL vs. SARSA & $5.79\times10^{-5}$ & 1.09 \\
& 6RX/2A & BGQL vs. Greedy & $2.37\times10^{-5}$ & 1.39 \\
& 6RX/2A & BGQL vs. Random & $4.86\times10^{-5}$ & 1.16 \\
\cline{2-5}
& 8RX/2A & BGQL vs. SARSA & $9.71\times10^{-5}$ & 0.95 \\
& 8RX/2A & BGQL vs. Greedy & $4.07\times10^{-5}$ & 1.23 \\
& 8RX/2A & BGQL vs. Random & $2.84\times10^{-5}$ & 1.07 \\
\hline

\shortstack[l]{Attacker-identification\\recall} & 4RX/1A & BGQL vs. SARSA & Not testable & Undefined \\
& 4RX/1A & BGQL vs. Greedy & $1.21\times10^{-6}$ & 2.86 \\
& 4RX/1A & BGQL vs. Random & $7.74\times10^{-6}$ & 1.39 \\
\cline{2-5}
& 6RX/1A & BGQL vs. SARSA & Not testable & Undefined \\
& 6RX/1A & BGQL vs. Greedy & $7.05\times10^{-7}$ & 2.98 \\
& 6RX/1A & BGQL vs. Random & $2.03\times10^{-7}$ & 2.95 \\
\cline{2-5}
& 6RX/2A & BGQL vs. SARSA & $1.00$ & 0.10 \\
& 6RX/2A & BGQL vs. Greedy & $5.37\times10^{-6}$ & 1.65 \\
& 6RX/2A & BGQL vs. Random & $3.96\times10^{-5}$ & 1.20 \\
\cline{2-5}
& 8RX/2A & BGQL vs. SARSA & $1.00$ & 0.18 \\
& 8RX/2A & BGQL vs. Greedy & $4.21\times10^{-6}$ & 1.77 \\
& 8RX/2A & BGQL vs. Random & $1.40\times10^{-5}$ & 1.44 \\

\hline
\end{tabular}
\end{table*}
\endgroup

The statistical analysis supports the performance trends in Figure~\ref{fig:Fig9}. Across all four evaluated configurations, BGQL achieves significantly higher aggregate full-band link-rate and rate-to-device-power scores than SARSA, Greedy, and Random, with large positive paired effect sizes. BGQL and SARSA achieve identical recall in the single-jammer configurations and show no statistically significant recall differences in the two-jammer configurations. BGQL also significantly outperforms Greedy and Random in attacker-identification recall. These results support the effectiveness of the proposed belief-guided beam-power control framework within the evaluated simulation setting.

\section{Conclusion}
\label{sec:conclusion}

This paper presented a simulation-based DT reference and BGQL scheme for
candidate-node jammer identification and joint beam-power control. In the four reported configurations, the complete BGQL implementation produced the
largest aggregate full-band link-rate and rate-to-device-power scores among
the evaluated methods. Its attacker detection accuracy was
$100\%$ in the one-jammer configurations and approximately $78\%$-$81\%$ in
the two-jammer configurations. These gains were accompanied by substantially
greater execution time because BGQL scores 625 candidate beam pairs during
exploratory steps. Our conclusions are limited to the reported site-specific, matched-model simulations. 
Moreover, the reported aggregate rate and rate-to-power quantities are comparison scores unless an explicit multiuser resource-sharing model is introduced. Future work will evaluate DT mismatch and measurement errors, unknown-cardinality jammer detection, and larger independent topologies.

\section*{CRediT authorship contribution statement}

Parmida Geranmayeh: Conceptualization, Methodology, Software,
Validation, Formal analysis, Investigation, Data curation,
Visualization, Writing - original draft, Writing - review and editing.

Onur G\"unl\"u: Conceptualization, Methodology, Supervision,
Project administration, Funding acquisition, Writing - review and editing.

\section*{Funding}

The work of OG was partially supported by the ZENITH Research and Leadership Career Development Fund under Grant ID23.01, EU COST Action 6G-PHYSEC,  Swedish Foundation for Strategic Research (SSF) under Grant ID24-0087, and German Federal Ministry of Research, Technology and Space (BMFTR) 6GEM+ Research Hub under the Grants 16KIS2412 and 16KISS005.

\section*{Declaration of competing interest}

The authors declare that they have no known competing financial
interests or personal relationships that could have appeared to
influence the work reported in this paper.

\section*{Data availability}

The data and code supporting the findings of this study are not publicly available.

\section*{Declaration of generative AI and AI-assisted
technologies in the manuscript preparation process}

During the preparation of this manuscript, the authors used
OpenAI ChatGPT to assist with language editing, clarification
of methodological descriptions, and improvement of presentation.
The authors reviewed and edited the resulting content and
take full responsibility for the content of the publication.

\bibliographystyle{IEEEtran}
\bibliography{References}

@ARTICLE{Fong2023Reliability,
  author = {Fong, B. and Kim, H. and Fong, A. C. M. and Hong, G. Y. and Tsang, K. F.},
  title = {Reliability optimization in the design and implementation of {6G} vehicle-to-infrastructure systems for emergency management in a smart city environment},
  journal = {IEEE Commun. Mag.},
  year = {2023},
  volume = {61},
  number = {8},
  pages = {148--153},
  doi = {10.1109/MCOM.002.2300840}
}

@MISC{Haider2020AI,
  author = {Haider, N. and Baig, M. Z. and Imran, M.},
  title = {Artificial intelligence and machine learning in {5G} network security: Opportunities, advantages, and future research trends},
  year = {2020},
  doi = {10.48550/arXiv.2007.04490},
  eprint = {2007.04490},
  archivePrefix = {arXiv},
  howpublished = {arXiv preprint arXiv:2007.04490}
}

@ARTICLE{Yu2022DeepLearning,
  author = {Yu, W. and Sohrabi, F. and Jiang, T.},
  title = {Role of deep learning in wireless communications},
  journal = {IEEE BITS Inf. Theory Mag.},
  year = {2022},
  volume = {2},
  number = {2},
  pages = {56--72},
  doi = {10.1109/MBITS.2022.3212978}
}

@ARTICLE{Bourtsoulatze2019DeepJSCC,
  author = {Bourtsoulatze, E. and Kurka, D. B. and G{\"u}nd{\"u}z, D.},
  title = {Deep joint source-channel coding for wireless image transmission},
  journal = {IEEE Trans. Cogn. Commun. Netw.},
  year = {2019},
  volume = {5},
  number = {3},
  pages = {567--579},
  doi = {10.1109/TCCN.2019.2919300}
}

@ARTICLE{Nachmani2018Decoding,
  author = {Nachmani, E. and Marciano, E. and Lugosch, L. and Gross, W. J. and Burshtein, D. and Be'ery, Y.},
  title = {Deep learning methods for improved decoding of linear codes},
  journal = {IEEE J. Sel. Topics Signal Process.},
  year = {2018},
  volume = {12},
  number = {1},
  pages = {119--131},
  doi = {10.1109/JSTSP.2017.2788405}
}

@ARTICLE{Kim2020PhysicalLayer,
  author = {Kim, H. and Oh, S. and Viswanath, P.},
  title = {Physical layer communication via deep learning},
  journal = {IEEE J. Sel. Areas Inf. Theory},
  year = {2020},
  volume = {1},
  number = {1},
  pages = {5--18},
  doi = {10.1109/JSAIT.2020.2991562}
}

@ARTICLE{Aoudia2022Waveform,
  author = {Ait Aoudia, F. and Hoydis, J.},
  title = {Waveform learning for next-generation wireless communication systems},
  journal = {IEEE Trans. Commun.},
  year = {2022},
  volume = {70},
  number = {6},
  pages = {3804--3817},
  doi = {10.1109/TCOMM.2022.3164060}
}

@ARTICLE{Ye2018OFDM,
  author = {Ye, H. and Li, G. Y. and Juang, B.-H.},
  title = {Power of deep learning for channel estimation and signal detection in {OFDM} systems},
  journal = {IEEE Wireless Commun. Lett.},
  year = {2018},
  volume = {7},
  number = {1},
  pages = {114--117},
  doi = {10.1109/LWC.2017.2757490}
}

@ARTICLE{Farsad2018Detection,
  author = {Farsad, N. and Goldsmith, A.},
  title = {Neural network detection of data sequences in communication systems},
  journal = {IEEE Trans. Signal Process.},
  year = {2018},
  volume = {66},
  number = {21},
  pages = {5663--5678},
  doi = {10.1109/TSP.2018.2868322}
}

@ARTICLE{Sohrabi2022Compression,
  author = {Sohrabi, F. and Jiang, T. and Yu, W.},
  title = {Learning progressive distributed compression strategies from local channel state information},
  journal = {IEEE J. Sel. Topics Signal Process.},
  year = {2022},
  volume = {16},
  number = {3},
  pages = {573--584},
  doi = {10.1109/JSTSP.2022.3158820}
}

@ARTICLE{Sun2018Optimize,
  author = {Sun, H. and Chen, X. and Shi, Q. and Hong, M. and Fu, X. and Sidiropoulos, N. D.},
  title = {Learning to optimize: Training deep neural networks for interference management},
  journal = {IEEE Trans. Signal Process.},
  year = {2018},
  volume = {66},
  number = {20},
  pages = {5438--5453},
  doi = {10.1109/TSP.2018.2866382}
}

@ARTICLE{Cui2019Spatial,
  author = {Cui, W. and Shen, K. and Yu, W.},
  title = {Spatial deep learning for wireless scheduling},
  journal = {IEEE J. Sel. Areas Commun.},
  year = {2019},
  volume = {37},
  number = {6},
  pages = {1248--1261}
}

@ARTICLE{Lee2021GraphEmbedding,
  author = {Lee, M. and Yu, G. and Li, G. Y.},
  title = {Graph embedding-based wireless link scheduling with few training samples},
  journal = {IEEE Trans. Wireless Commun.},
  year = {2021},
  volume = {20},
  number = {4},
  pages = {2282--2294},
  doi = {10.1109/TWC.2020.3040983}
}

@ARTICLE{Shen2021GNN,
  author = {Shen, Y. and Shi, Y. and Zhang, J. and Letaief, K. B.},
  title = {Graph neural networks for scalable radio resource management: Architecture design and theoretical analysis},
  journal = {IEEE J. Sel. Areas Commun.},
  year = {2021},
  volume = {39},
  number = {1},
  pages = {101--115},
  doi = {10.1109/JSAC.2020.3036965}
}

@ARTICLE{Sohrabi2021Beamforming,
  author = {Sohrabi, F. and Chen, Z. and Yu, W.},
  title = {Deep active learning approach to adaptive beamforming for {mmWave} initial alignment},
  journal = {IEEE J. Sel. Areas Commun.},
  year = {2021},
  volume = {39},
  number = {8},
  pages = {2347--2360},
  doi = {10.1109/JSAC.2021.3087234}
}

@ARTICLE{Zhang2022RIS,
  author = {Zhang, Z. and Jiang, T. and Yu, W.},
  title = {Learning-based user scheduling in reconfigurable intelligent surface assisted multiuser downlink},
  journal = {IEEE J. Sel. Topics Signal Process.},
  year = {2022},
  volume = {16},
  number = {5},
  pages = {1026--1039},
  doi = {10.1109/JSTSP.2022.3178213}
}

@ARTICLE{He2018Channel,
  author = {He, H. and Wen, C.-K. and Jin, S. and Li, G. Y.},
  title = {Deep learning-based channel estimation for beamspace {mmWave} massive {MIMO} systems},
  journal = {IEEE Wireless Commun. Lett.},
  year = {2018},
  volume = {7},
  number = {5},
  pages = {852--855},
  doi = {10.1109/LWC.2018.2832128}
}

@ARTICLE{Neumann2018MMSE,
  author = {Neumann, D. and Wiese, T. and Utschick, W.},
  title = {Learning the {MMSE} channel estimator},
  journal = {IEEE Trans. Signal Process.},
  year = {2018},
  volume = {66},
  number = {11},
  pages = {2905--2917},
  doi = {10.1109/TSP.2018.2799164}
}

@ARTICLE{Ali2021MassiveIoT,
  author = {Ali, R. and Ashraf, I. and Bashir, A. K. and Zikria, Y. B.},
  title = {Reinforcement-learning-enabled massive Internet of Things for {6G} wireless communications},
  journal = {IEEE Commun. Standards Mag.},
  year = {2021},
  volume = {5},
  number = {2},
  pages = {126--131},
  doi = {10.1109/MCOMSTD.001.2000055}
}

@ARTICLE{Wang2020Channels,
  author = {Wang, C.-X. and Huang, J. and Wang, H. and Gao, X. and You, X. and Hao, Y.},
  title = {{6G} wireless channel measurements and models: Trends and challenges},
  journal = {IEEE Veh. Technol. Mag.},
  year = {2020},
  volume = {15},
  number = {4},
  pages = {22--32},
  doi = {10.1109/MVT.2020.3018436}
}

@ARTICLE{Demirhan2023ISAC,
  author = {Demirhan, U. and Alkhateeb, A.},
  title = {Integrated sensing and communication for {6G}: Ten key machine learning roles},
  journal = {IEEE Commun. Mag.},
  year = {2023},
  volume = {61},
  number = {5},
  pages = {113--119},
  doi = {10.1109/MCOM.006.2200480}
}

@ARTICLE{Salami2023Radar,
  author = {Salami, D. and Ning, W. and Ruttik, K. and J{\"a}ntti, R. and Sigg, S.},
  title = {A joint radar and communication approach for {5G NR} using reinforcement learning},
  journal = {IEEE Commun. Mag.},
  year = {2023},
  volume = {61},
  number = {5},
  pages = {106--112},
  doi = {10.1109/MCOM.001.2200474}
}

@ARTICLE{Ayed2023Accordion,
  author = {Ayed, F. and De Domenico, A. and Garcia-Rodriguez, A. and L{\'o}pez-P{\'e}rez, D.},
  title = {Accordion: A communication-aware machine learning framework for next generation networks},
  journal = {IEEE Commun. Mag.},
  year = {2023},
  volume = {61},
  number = {6},
  pages = {104--110},
  doi = {10.1109/MCOM.001.2200358}
}

@ARTICLE{Ferreira2019Satellite,
  author = {Ferreira, P. V. R. and Paffenroth, R. and Wyglinski, A. M. and Hackett, T. M. and Bilen, S. G. and Reinhart, R. C. and Mortensen, D. J.},
  title = {Reinforcement learning for satellite communications: From {LEO} to deep space operations},
  journal = {IEEE Commun. Mag.},
  year = {2019},
  volume = {57},
  number = {5},
  pages = {70--75},
  doi = {10.1109/MCOM.2019.1800796}
}

@ARTICLE{Qi2021UAVBeamforming,
  author = {Qi, F. and Li, W. and Yu, P. and Feng, L. and Zhou, F.},
  title = {Deep learning-based {BackCom} multiple beamforming for {6G} {UAV} {IoT} networks},
  journal = {EURASIP J. Wireless Commun. Netw.},
  year = {2021},
  volume = {2021},
  pages = {50},
  doi = {10.1186/s13638-021-01932-4}
}

@ARTICLE{Ihsan2022IRS,
  author = {Ihsan, A. and Chen, W. and Asif, M. and Khan, W. U. and Wu, Q. and Li, J.},
  title = {Energy-efficient {IRS}-aided {NOMA} beamforming for {6G} wireless communications},
  journal = {IEEE Trans. Green Commun. Netw.},
  year = {2022},
  volume = {6},
  number = {4},
  pages = {1945--1956},
  doi = {10.1109/TGCN.2022.3209617}
}

@INPROCEEDINGS{Jiang2022ICC,
  author = {Jiang, W. and Schotten, H. D.},
  title = {Initial access for millimeter-wave and terahertz communications with hybrid beamforming},
  booktitle = {Proc. IEEE Int. Conf. Commun. (ICC)},
  year = {2022},
  pages = {3960--3965},
  doi = {10.1109/ICC45855.2022.9838386}
}

@ARTICLE{Ding2022DigitalTwin,
  author = {Ding, C. and Ho, I. W.-H.},
  title = {Digital-twin-enabled city-model-aware deep learning for dynamic channel estimation in urban vehicular environments},
  journal = {IEEE Trans. Green Commun. Netw.},
  year = {2022},
  volume = {6},
  number = {3},
  pages = {1604--1612},
  doi = {10.1109/TGCN.2022.3173414}
}

@ARTICLE{Geok2018RayTracing,
  author = {Geok, T. K. and Hossain, F. and Kamaruddin, M. N. and Rahman, N. Z. A. and Thiagarajah, S. and Chiat, A. T. W. and Hossen, J. and Liew, C. P.},
  title = {A comprehensive review of efficient ray-tracing techniques for wireless communication},
  journal = {Int. J. Commun. Antenna Propag.},
  year = {2018},
  volume = {8},
  number = {2},
  pages = {123--136},
  doi = {10.15866/irecap.v8i2.13797}
}

@ARTICLE{OnurRoleofISAC,
  author = {Qaisar, M. U. F. and Yuan, W. and G{\"u}nl{\"u}, O. and Riihonen, T. and Cui, Y. and Zhang, L. and Gonzalez-Prelcic, N. and Di Renzo, M. and Han, Z.},
  title = {The role of {ISAC} in {6G} networks: {Enabling} next-generation wireless systems},
  journal = {IEEE Trans. Netw. Sci. Eng.},
  year = {2026},
  volume = {13},
  pages = {7825--7861},
  doi = {10.1109/TNSE.2026.3666665}
}

@ARTICLE{OnurFurkanISACTutorial,
  author = {Keskin, M. F. and Srinivasan, M. and G{\"u}nl{\"u}, O. and Chen, H. and Papadimitratos, P. and Almgren, M. and He, Z. S. and Wymeersch, H.},
  title = {Multi-domain security for {6G ISAC}: Challenges and opportunities in transportation},
  journal = {IEEE Commun. Mag.},
  year = {2026},
  volume = {64},
  number = {9},
  pages = {70--77},
  doi = {10.1109/MCOM.001.2500712}
}

@MISC{Sheen2020DigitalTwinRIS,
  author = {Sheen, B. and Yang, J. and Feng, X. and Chowdhury, M.},
  title = {A digital twin for reconfigurable intelligent surface assisted wireless communication},
  year = {2020},
  doi = {10.48550/arXiv.2009.00454},
  eprint = {2009.00454},
  archivePrefix = {arXiv},
  howpublished = {arXiv preprint arXiv:2009.00454}
}

@INPROCEEDINGS{Jalali2023SWIPT,
  author = {Jalali, J. and Khalili, A. and Rezaei, A. and Famaey, J. and Saad, W.},
  title = {Power-efficient antenna switching and beamforming design for multi-user {SWIPT} with non-linear energy harvesting},
  booktitle = {Proc. IEEE 20th Consum. Commun. Netw. Conf. (CCNC)},
  year = {2023},
  pages = {746--751},
  doi = {10.1109/CCNC51644.2023.10059879}
}

@ARTICLE{Bakhtiar2023Beamforming,
  author = {Karim, B. A. and Ali, H. K.},
  title = {A novel beamforming technique using {mmWave} antenna arrays for {5G} wireless communication networks},
  journal = {Digit. Signal Process.},
  year = {2023},
  volume = {134},
  pages = {103917},
  doi = {10.1016/j.dsp.2023.103917}
}

@ARTICLE{Chary2024AIBeamforming,
  author = {Chary, M. K. and Krishna, C. H. V. and Krishna, D. R.},
  title = {Accurate channel estimation and hybrid beamforming using artificial intelligence for massive {MIMO} {5G} systems},
  journal = {AEU--Int. J. Electron. Commun.},
  year = {2024},
  volume = {173},
  pages = {154971},
  doi = {10.1016/j.aeue.2023.154971}
}

@INCOLLECTION{Geranmayeh2022BeamSelection,
  author = {Geranmayeh, P. and Sedunova, E. and Grass, E.},
  title = {Efficient beam selection for increased overall wireless network capacity},
  booktitle = {Recent Advances in Communication Networks and Embedded Systems},
  publisher = {Springer},
  address = {Cham},
  year = {2024},
  pages = {14--27},
  doi = {10.1007/978-3-031-59619-3_2}
}

@ARTICLE{Ji2021RayTracingIndoor,
  author = {Ji, Z. and Li, B.-H. and Wang, H.-X. and Chen, H.-Y. and Sarkar, T. K.},
  title = {Efficient ray-tracing methods for propagation prediction for indoor wireless communications},
  journal = {IEEE Antennas Propag. Mag.},
  year = {2001},
  volume = {43},
  number = {2},
  pages = {41--49},
  doi = {10.1109/74.924603}
}

@INPROCEEDINGS{Hsiao2017RayTracing,
  author = {Hsiao, A.-Y. and Yang, C.-F. and Wang, T.-S. and Lin, I. and Liao, W.-J.},
  title = {Ray-tracing simulations for millimeter wave propagation in {5G} wireless communications},
  booktitle = {Proc. IEEE Int. Symp. Antennas Propag. USNC/URSI Natl. Radio Sci. Meeting},
  year = {2017},
  pages = {1901--1902},
  doi = {10.1109/APUSNCURSINRSM.2017.8072993}
}

@ARTICLE{Yun2015RayTracingTheory,
  author = {Yun, Z. and Iskander, M. F.},
  title = {Ray tracing for radio propagation modeling: Principles and applications},
  journal = {IEEE Access},
  year = {2015},
  volume = {3},
  pages = {1089--1100},
  doi = {10.1109/ACCESS.2015.2453991}
}

@ARTICLE{Tiberi2009UWB,
  author = {Tiberi, G. and Bertini, S. and Malik, W. Q. and Monorchio, A. and Edwards, D. J. and Manara, G.},
  title = {Analysis of realistic ultrawideband indoor communication channels by using an efficient ray-tracing based method},
  journal = {IEEE Trans. Antennas Propag.},
  year = {2009},
  volume = {57},
  number = {3},
  pages = {777--785},
  doi = {10.1109/TAP.2009.2013432}
}

@ARTICLE{OnurSecureISACJournal,
  author = {G{\"u}nl{\"u}, O. and Bloch, M. R. and Schaefer, R. F. and Yener, A.},
  title = {Secure integrated sensing and communication},
  journal = {IEEE J. Sel. Areas Inf. Theory},
  year = {2023},
  volume = {4},
  pages = {40--53},
  doi = {10.1109/JSAIT.2023.3275048}
}

@ARTICLE{Onur2026secureISACTutorial,
  author = {Welling, T. and G{\"u}nl{\"u}, O. and Yener, A.},
  title = {Secure integrated sensing and communication: Information theory offers insights},
  journal = {IEEE BITS Inf. Theory Mag.},
  year = {2026},
  doi = {10.1109/MBITS.2026.3728769},
  note = {accepted}
}

@ARTICLE{OnurSecurityTutorial,
  author = {Bloch, M. R. and G{\"u}nl{\"u}, O. and Yener, A. and Oggier, F. E. and Poor, H. V. and Sankar, L. and Schaefer, R. F.},
  title = {An overview of information-theoretic security and privacy: Metrics, limits and applications},
  journal = {IEEE J. Sel. Areas Inf. Theory},
  year = {2021},
  volume = {2},
  number = {1},
  pages = {5--22},
  doi = {10.1109/JSAIT.2021.3062755}
}

@INPROCEEDINGS{Zhou2026FeedbackLunch,
  author = {Zhou, Y. and Devroye, N. and G{\"u}nl{\"u}, O.},
  title = {Feedback lunch: Learned feedback codes for secure communications},
  booktitle = {Proc. ACM Workshop Wireless Security Mach. Learn. (WiseML)},
  year = {2026},
  pages = {19--24},
  doi = {10.1145/3811880.3815100}
}

@INPROCEEDINGS{Yu2023Covert,
  author = {Yu, F. and Liu, C. and Shi, Z. and Lin, M. and Zheng, T.-X. and Yao, B.},
  title = {Covert communication assisted by friendly jammer with jamming-monitoring channel uncertainty},
  booktitle = {Proc. 9th Int. Conf. Mech. Electron. Eng. (ICMEE)},
  year = {2023},
  pages = {43--48},
  doi = {10.1109/ICMEE59781.2023.10525396}
}

@ARTICLE{Esmaili2026Covert,
  author = {Esmaili, E. and Hajizadeh, R. and Forouzesh, M.},
  title = {Machine learning based detection of covert communications under jamming interference},
  journal = {Sci. Rep.},
  year = {2026},
  volume = {16},
  pages = {23003},
  doi = {10.1038/s41598-026-53830-8}
}

@ARTICLE{Geranmayeh2025DQN,
  author = {Geranmayeh, P. and Grass, E.},
  title = {Optimization of beamforming and transmit power using {DQN} and comparison with traditional techniques},
  journal = {IEEE Access},
  year = {2025},
  volume = {13},
  pages = {94275--94285},
  doi = {10.1109/ACCESS.2025.3573096}
}

@INPROCEEDINGS{Vivekanand2024DRL,
  author = {Vivekanand, C. V. and Talele, G. and Jaishanthi, B. and Gopalan, S. H. and Maheswari, K. and Maheswari, B. U.},
  title = {Deep reinforcement learning for resource allocation in wireless communication systems},
  booktitle = {Proc. 9th Int. Conf. Sci. Technol. Eng. Math. (ICONSTEM)},
  year = {2024},
  pages = {1--5},
  doi = {10.1109/ICONSTEM60960.2024.10568620}
}

@ARTICLE{Elsadig2022Covert,
  author = {Elsadig, M. A. and Gafar, A.},
  title = {Covert channel detection: Machine learning approaches},
  journal = {IEEE Access},
  year = {2022},
  volume = {10},
  pages = {38391--38405},
  doi = {10.1109/ACCESS.2022.3164392}
}

@ARTICLE{Jere2023BayesianJamming,
  author = {Jere, S. and Wang, Y. and Aryendu, I. and Dayekh, S. and Liu, L.},
  title = {{Bayesian} inference-assisted machine learning for near real-time jamming detection and classification in {5G} New Radio ({NR})},
  journal = {IEEE Trans. Wireless Commun.},
  year = {2024},
  volume = {23},
  number = {7},
  pages = {7043--7059},
  doi = {10.1109/TWC.2023.3337058}
}

@INPROCEEDINGS{Huang2022VANET,
  author = {Huang, F. and Li, Q. and Zhao, J.},
  title = {Trust management model of {VANETs} based on machine learning and active detection technology},
  booktitle = {Proc. IEEE/CIC Int. Conf. Commun. China Workshops},
  year = {2022},
  pages = {412--416},
  doi = {10.1109/ICCCWorkshops55477.2022.9896700}
}

@ARTICLE{Lei2018Game,
  author = {Lei, C. and Zhang, H. Q. and Wan, L. M. and Liu, L. and Ma, D. H.},
  title = {Incomplete information {Markov} game theoretic approach to strategy generation for moving target defense},
  journal = {Comput. Commun.},
  year = {2018},
  volume = {116},
  pages = {184--199},
  doi = {10.1016/j.comcom.2017.12.001}
}

@ARTICLE{Jiang2021Game,
  author = {Jiang, L{\"u} and Zhang, Hengwei and Wang, Jindong},
  title = {A {Markov} signaling game-theoretic approach to moving target defense strategy selection},
  journal = {Acta Electronica Sinica},
  year = {2021},
  volume = {49},
  number = {3},
  pages = {527--535},
  doi = {10.12263/DZXB.20191070}
}

@INPROCEEDINGS{Teofilo2012Poker,
  author = {Te{\'o}filo, L. F. and Passos, N. and Reis, L. P. and Cardoso, H. L.},
  title = {Adapting strategies to opponent models in incomplete information games: A reinforcement learning approach for poker},
  booktitle = {Proc. Int. Conf. Auton. Intell. Syst.},
  year = {2012},
  pages = {220--227},
  doi = {10.1007/978-3-642-31368-4_26}
}

@ARTICLE{Ren2023UAV,
  author = {Ren, Z. and Zhang, D. and Tang, S. and Xiong, W. and Yang, S.-H.},
  title = {Cooperative maneuver decision making for multi-{UAV} air combat based on incomplete information dynamic game},
  journal = {Def. Technol.},
  year = {2023},
  volume = {27},
  pages = {308--317},
  doi = {10.1016/j.dt.2022.10.008}
}

@ARTICLE{Yao2024BayesianGame,
  author = {Yao, P. and Jiang, Z. and Yan, B. and Yang, Q. and Wang, W.},
  title = {{Bayesian} and stochastic game joint approach for cross-layer optimal defensive decision-making in industrial cyber-physical systems},
  journal = {Inf. Sci.},
  year = {2024},
  volume = {662},
  pages = {120216},
  doi = {10.1016/j.ins.2024.120216}
}

@ARTICLE{Mamaghani2026ISAC,
  author = {Mamaghani, M. T. and Zhou, X. and Yang, N. and Swindlehurst, A. L.},
  title = {Securing integrated sensing and communication against a mobile adversary: A {Stackelberg} game with deep reinforcement learning},
  journal = {IEEE J. Sel. Areas Commun.},
  year = {2026},
  volume = {44},
  number = {4},
  pages = {942--958},
  doi = {10.1109/JSAC.2025.3611404}
}

@ARTICLE{Ruah2023BayesianDT,
  author = {Ruah, C. and Simeone, O. and Al-Hashimi, B.},
  title = {A {Bayesian} framework for digital twin-based control, monitoring, and data collection in wireless systems},
  journal = {IEEE J. Sel. Areas Commun.},
  year = {2023},
  volume = {41},
  number = {10},
  pages = {3146--3160},
  doi = {10.1109/JSAC.2023.3310093}
}

@INPROCEEDINGS{Krause2023DTAnomaly,
  author = {Krause, A. and Khursheed, M. D. and Schulz, P. and Burmeister, F. and Fettweis, G. P.},
  title = {Digital twin of the radio environment: A novel approach for anomaly detection in wireless networks},
  booktitle = {Proc. IEEE Globecom Workshops (GC Wkshps)},
  year = {2023},
  pages = {1307--1312},
  doi = {10.1109/GCWkshps58843.2023.10464447}
}

@INPROCEEDINGS{Geranmayeh2026ISAC,
  author = {Geranmayeh, P. and G{\"u}nl{\"u}, O.},
  title = {{6G} sensing security: Distributed game-theoretic {RL} for urban beamforming and attacker detection},
  booktitle = {Proc. IEEE Conference on Integrated Sensing and Communications (ISAC)},
  year = {2026},
  note = {accepted}
}

@article{Geranmayeh2026Bayesian,
  author  = {Geranmayeh, Parmida and G{\"u}nl{\"u}, Onur},
  title   = {{Bayesian}-Guided Cooperative {RL} Beamforming for Wireless Adversarial User Detection},
  journal = {Entropy},
  year    = {2026},
  volume  = {28},
  number  = {10},
  pages   = {1070},
  doi     = {10.3390/e28101070},
  url     = {https://www.mdpi.com/1099-4300/28/10/1070}
}
\begin{IEEEbiography}
[{\includegraphics[width=1in,height=1.25in,clip,keepaspectratio]{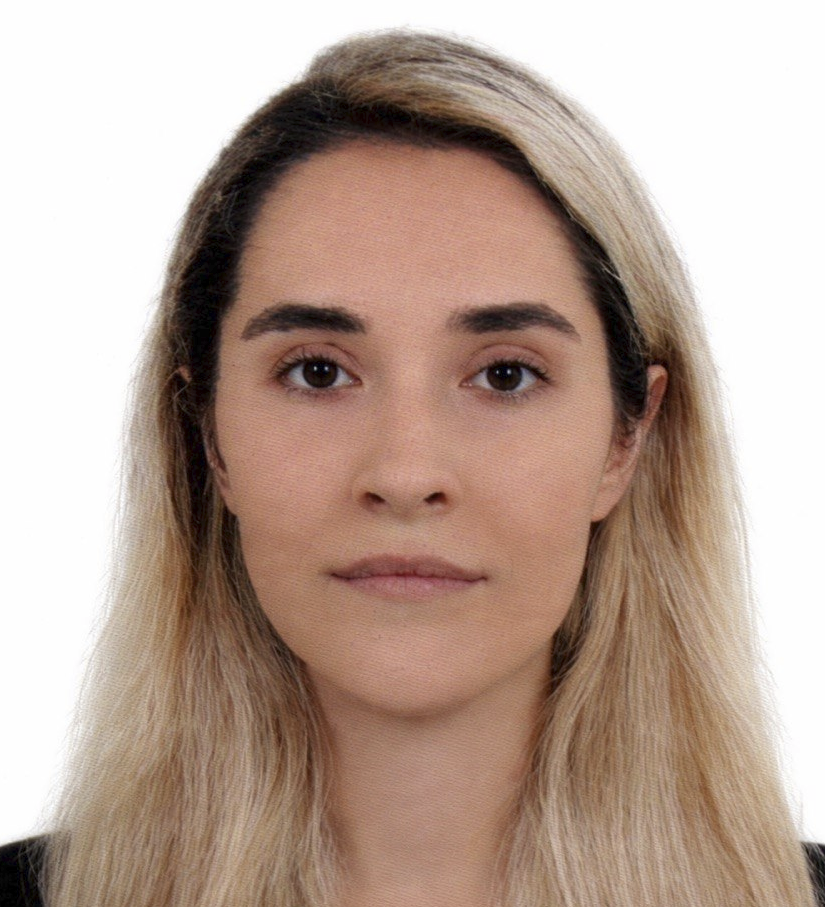}}]
{Parmida Geranmayeh}
received the B.Sc. degree in Computer Engineering (Software Engineering)
and the M.Sc. degree in Information Technology Engineering from the Iran
University of Science and Technology (IUST), Tehran, Iran, in 2018 and
2020, respectively. She received the Dr.-Ing. degree in Computer
Engineering from Humboldt University of Berlin, Germany, in 2025.
She is currently a Postdoctoral Researcher with the Department of
Communications Engineering (Nachrichtentechnik), Technical University
of Dortmund, Germany. Her research interests include machine learning,
reinforcement learning, optimization, beamforming, and 5G/6G wireless
communication systems. She serves as a reviewer for several international
journals, including \emph{Physical Communication} (Elsevier),
\emph{Scientific Reports} (Springer Nature), and the
\emph{International Journal of Communication Systems} (Wiley).
She served as a Session Chair at the IEEE-APS Topical Conference on
Antennas and Propagation in Wireless Communications in 2024 and received
the Best Presenter Award at the IEEE International Conference on
Artificial Intelligence for Sustainable Innovation (AI-SI 2025).
\end{IEEEbiography}

\begin{IEEEbiography}
[{\includegraphics[width=1in,height=1.25in,clip,keepaspectratio]{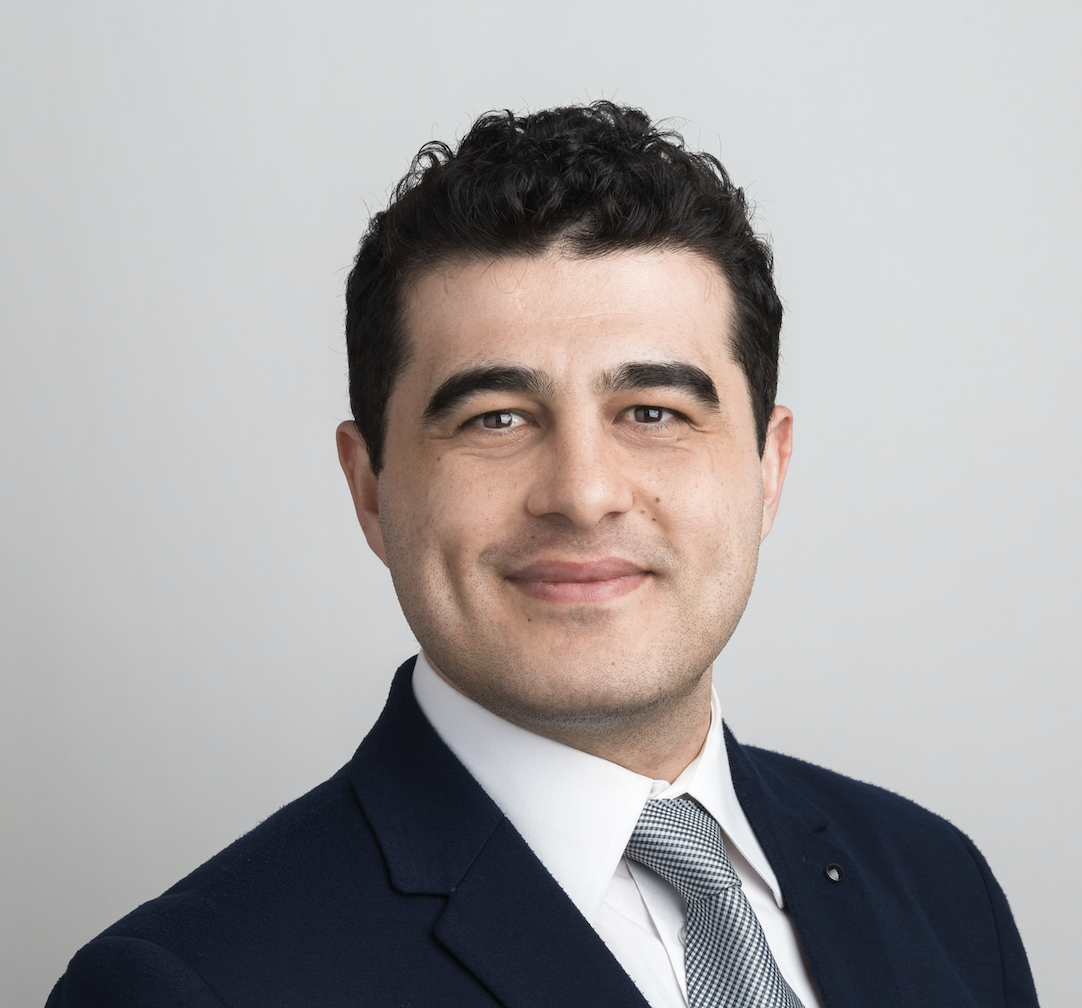}}]
{Onur G{\"u}nl{\"u}}
received the B.Sc. degree (Highest Distinction) in Electrical and
Electronics Engineering from Bilkent University, Turkey in 2011;
M.Sc. (Highest Distinction) and Dr.-Ing. (Ph.D. equivalent) degrees in
Communications Engineering both from the Technical University of Munich
(TUM), Germany in 2013 and 2018, respectively.
He was a Working Student in the Communication Systems division of Intel
Mobile Communications (IMC), now Apple Inc., in Munich, Germany during
November 2012--March 2013. Onur worked as a Research and Teaching
Assistant at TUM Chair of Communications Engineering (LNT) between
February 2014--May 2019. As a Visiting Researcher, among more than twenty
Research Stays at Top Universities and Companies, he was at TU Eindhoven,
Netherlands during February 2018--March 2018.
Onur was a Visiting Research Group Leader at Georgia Institute of
Technology, Atlanta, USA during February 2022--March 2022.
He was also a Visiting Professor at TU Dresden, Germany during
February 2023--March 2023.
Following Research Associate and Group Leader positions at TUM,
TU Berlin, and the University of Siegen, he joined Link{\"o}ping
University in October 2022 as an ELLIIT Assistant Professor and obtained
tenure as an Associate Professor leading the Information Theory and
Security Laboratory (ITSL) in August 2024.
He became a Swedish Docent (Dr.-habil.) of Information Theory in
December 2023 and an IEEE Senior Member in July 2024.
Since September 2025, Onur has been a Tenured Full Professor leading the
Lehrstuhl f{\"u}r Nachrichtentechnik (Institute of Communications
Engineering) at TU Dortmund (TUDO), Germany.
He was also a Guest Professor at Link{\"o}ping University, Sweden and
since September 2026, he is a Guest Full Professor there.
He has received the 2026 IEEE Communications Society \& Information
Theory Society---Joint Best Paper Award, 2025 IEEE Information Theory
Society---Joy Thomas Tutorial Paper Award, the 2023 ZENITH Research and
Career Development Award, 2021 IEEE Transactions on
Communications---Exemplary Reviewer Award, and the prestigious 2021 VDE
Information Technology Society (ITG) Johann-Philipp-Reis Award.
His research interests include distributed function computation,
information-theoretic privacy and security, coding theory, 6G integrated
sensing and communication (ISAC), and private learning.
Among his publications is the book
\emph{Key Agreement with Physical Unclonable Functions and Biometric
Identifiers} (Dr. Hut Verlag, 2019).
He serves as Associate Editor for
{IEEE JOURNAL ON SELECTED AREAS IN COMMUNICATIONS},
{IEEE TRANSACTIONS ON COMMUNICATIONS}, and {ENTROPY} Journal,
and was recently an Associate Editor of
{EURASIP JOURNAL ON WIRELESS COMMUNICATIONS AND NETWORKING}
and a Guest Editor of
{IEEE JOURNAL ON SELECTED AREAS IN INFORMATION THEORY}.
He also serves as a Working Group Leader for EU COST Action 6G-PHYSEC
on Intelligent and Resilient Systems and as an Elected Member of the
IEEE Signal Processing Society---Information Theory and Forensics
(IFS) Technical Committee (TC).
\end{IEEEbiography}

\EOD
\end{document}